\documentclass{WileyMSP-template}
\usepackage{makecell,color}
\usepackage{graphicx} 
\usepackage{algpseudocode}
\usepackage{algorithm}
\usepackage{graphicx} 
\usepackage{hyperref}
\usepackage{amsmath}
\usepackage{amsthm}
\usepackage{amssymb}
\usepackage{booktabs}
\usepackage{multirow}
\usepackage{geometry}
\usepackage{caption}
\begin{document}

\pagestyle{fancy}
\rhead{\includegraphics[width=2.5cm]{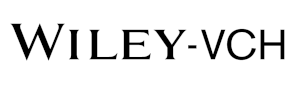}}

\title{A Systematic Computational Framework for Practical Identifiability Analysis in Mathematical Models Arising from Biology}

\maketitle


\author{Shun Wang}
\author{Wenrui Hao*}

\begin{affiliations}
 Department of Mathematics, Penn State University, University Park, Pennsylvania, United States of America\\
Email: wxh64@psu.edu

\end{affiliations}



\begin{abstract}
Practical identifiability is a fundamental challenge in the data-driven modeling of biological systems, as many model parameters cannot be directly measured and must be estimated from experimental data. Without confirming the identifiability of these parameters, model predictions may be unreliable, limiting their usefulness for understanding biological mechanisms or informing experimental and clinical decisions. In this paper, we propose a novel mathematical framework for practical identifiability analysis in dynamic models. Starting from a rigorous mathematical definition, we prove that practical identifiability is equivalent to the invertibility of the Fisher Information Matrix (FIM). We further establish the relationship between practical identifiability and coordinate identifiability, introducing an efficient metric that simplifies and accelerates identifiability assessment compared to traditional profile likelihood methods. To address non-identifiable parameters, we incorporate new regularization terms, enabling uncertainty quantification and improving model reliability. Additionally, we develop an optimal experimental design algorithm to ensure all parameters are practically identifiable from collected data. Applications to Hill functions, neural networks, and biological models demonstrate the effectiveness and computational efficiency of the proposed framework in uncovering critical biological processes and identifying key observable variables.

\end{abstract}


\section*{Introduction}
In systems biology, mathematical modeling is a widely used and powerful tool for analyzing biological processes across multiple scales. At the microscopic scale, differential equations are employed  to model intracellular signaling networks \cite{alon2019introduction,qiao2019network,ma2009defining}, including cancer signaling pathways \cite{aguda2008microrna}, epithelial-mesenchymal transitions \cite{lang2021landscape}, single-cell RNA velocity \cite{su2024hodge,la2018rna,sha2024reconstructing}, and morphogen gradients involved in cell development \cite{lander2002morphogen,shen2022scaling,rodriguez2022concentration,wei2016dual}. At the mesoscopic scale, ordinary differential equations (ODEs) are frequently applied to simulate cancer-immune \cite{kreger2023myeloid,anderson2024global,liao2022mathematical} and virus-host immune interactions \cite{zhou2023dynamical,perelson2002modelling}, aiding in the prediction of disease progression. At the macroscopic scale, partial differential equations (PDEs) are employed to describe cell movement and spatial cell-cell interactions, such as tumor cell invasion \cite{eisenberg2011mechanistic,stepien2018traveling,kim2023role} and spatial interactions of immune cells \cite{mirzaei2023investigating,lai2018modeling}, facilitating predictions of cancer development and cardiovascular disease progression \cite{davey2024simulating,peskin1977numerical,peskin2020cardiac}.

Due to technical constraints and other limitations, not all parameters in these models can be directly observed. To accurately reflect real-world dynamics, it is essential to calibrate model parameters using observable data. Typically, the least squares method is employed to estimate unmeasured model parameters based on observable data \cite{gabor2015robust}. However, there may be cases where certain unknown parameters are inherently non-identifiable from the observable data, while others exhibit high sensitivity to it. Such situations can result in different parameter sets producing similar dynamic trajectories, raising significant concerns about the reliability and accuracy of the model's predictions. Consequently, parameter identifiability has become a critical issue in the development and application of mathematical models \cite{gallo2022lack}.

Parameter identifiability consists of two components: structural identifiability and practical identifiability \cite{wieland2021structural}. Structural identifiability, or prior identifiability, is defined as the condition in which two sets of observed variables or system outputs are identical if and only if their corresponding parameter sets are exactly the same \cite{miao2011identifiability}. The primary goal of structural identifiability analysis is to determine whether a model is theoretically identifiable by examining its structure before attempting to estimate parameters from data. Several computational methods have been developed for structural identifiability analysis, with differential algebra \cite{miao2011identifiability} and Lie derivatives \cite{villaverde2016structural} being among the most widely  used approaches. Furthermore, various software tools have been designed for structural identifiability analysis of dynamic systems, such as GenSSI2 \cite{ligon2018genssi}, SIAN \cite{hong2019sian}, and STIKE-GOLDD \cite{villaverde2016structural}. These tools have been benchmarked against standard models to assess and compare their performance \cite{rey2023benchmarking}. However, structural identifiability analysis relies on two key assumptions: that model structures are entirely accurate and that measurements are error-free \cite{miao2011identifiability}. Since these assumptions rarely hold in practice, it is essential to determine whether structurally identifiable parameters can be reliably estimated from noisy, imperfect  data. As such, only models determined to be structurally identifiable proceed to practical identifiability analysis \cite{miao2011identifiability}.

Practical identifiability, or posterior  identifiability,  refers to the ability to assess parameter identifiability based on actual experimental data \cite{miao2011identifiability}. Unlike structural identifiability, practical identifiability lacks a rigorous, universally accepted mathematical definition — an issue that remains open and urgently needs addressing. Nevertheless, compared to structural identifiability, practical identifiability offers greater direct relevance for applied modeling.
For instance, one study employed the Hessian matrix to evaluate the practical identifiability of observable and hidden variables in models, enabling the quantification of uncertainties associated with unobservable variables \cite{gallo2022lack}. Additionally, another study utilized non-identifiable parameters to analyze parameter uncertainty when mathematical models were fitted to data \cite{monsalve2022analysis}. Furthermore, practical identifiability has been applied to design minimally sufficient experiments for pharmacokinetic/pharmacodynamic models that capture the distribution of drugs within the tumor microenvironment \cite{gevertz2024minimally}. Typically, practical identifiability is evaluated using methods such as calculating the profile likelihood \cite{wieland2021structural,raue2009structural,raue2013joining,ciocanel2024parameter} or the parameter correlation matrix through the FIM \cite{miao2011identifiability,komorowski2011sensitivity,joubert2018determining}. However, calculating the profile likelihood is computationally expensive, particularly when the number of model parameters is large. Meanwhile, the FIM-based approach is limited to cases where the FIM is invertible, as all the parameters are practically identifiable if and only if the FIM is invertible \cite{miao2011identifiability,komorowski2011sensitivity}. Addressing practical identifiability when the FIM is singular remains one of the central challenges in this field.

In this paper, we propose a novel and rigorous mathematical definition for practical identifiability, proving that the invertibility of the FIM is a necessary and sufficient condition for all the parameters to be practically identifiable. Using the concept of coordinate identifiability derived from the profile likelihood \cite{raue2009structural}, we establish the relationship between practical identifiability and coordinate identifiability and introduce a more effective metric for analyzing parameter coordinate identifiability. To address cases where the FIM is singular, we identify the eigenvectors associated with non-identifiable parameters through eigenvalue decomposition (EVD) and incorporate these eigenvectors into practical identifiability and regularization terms, enabling all the parameters to become practically identifiable during model fitting. Additionally, we develop an uncertainty quantification method to assess the influence of non-identifiable parameters on model predictions. Finally, we propose a novel algorithm for experiment design that ensures the collected data can render all model parameters practically identifiable.

\section*{Results}
\subsection*{Overview of Practical Identifiability Analysis and Its Applications}
\begin{figure}[ht]
\centering
\includegraphics[width=0.9\linewidth]{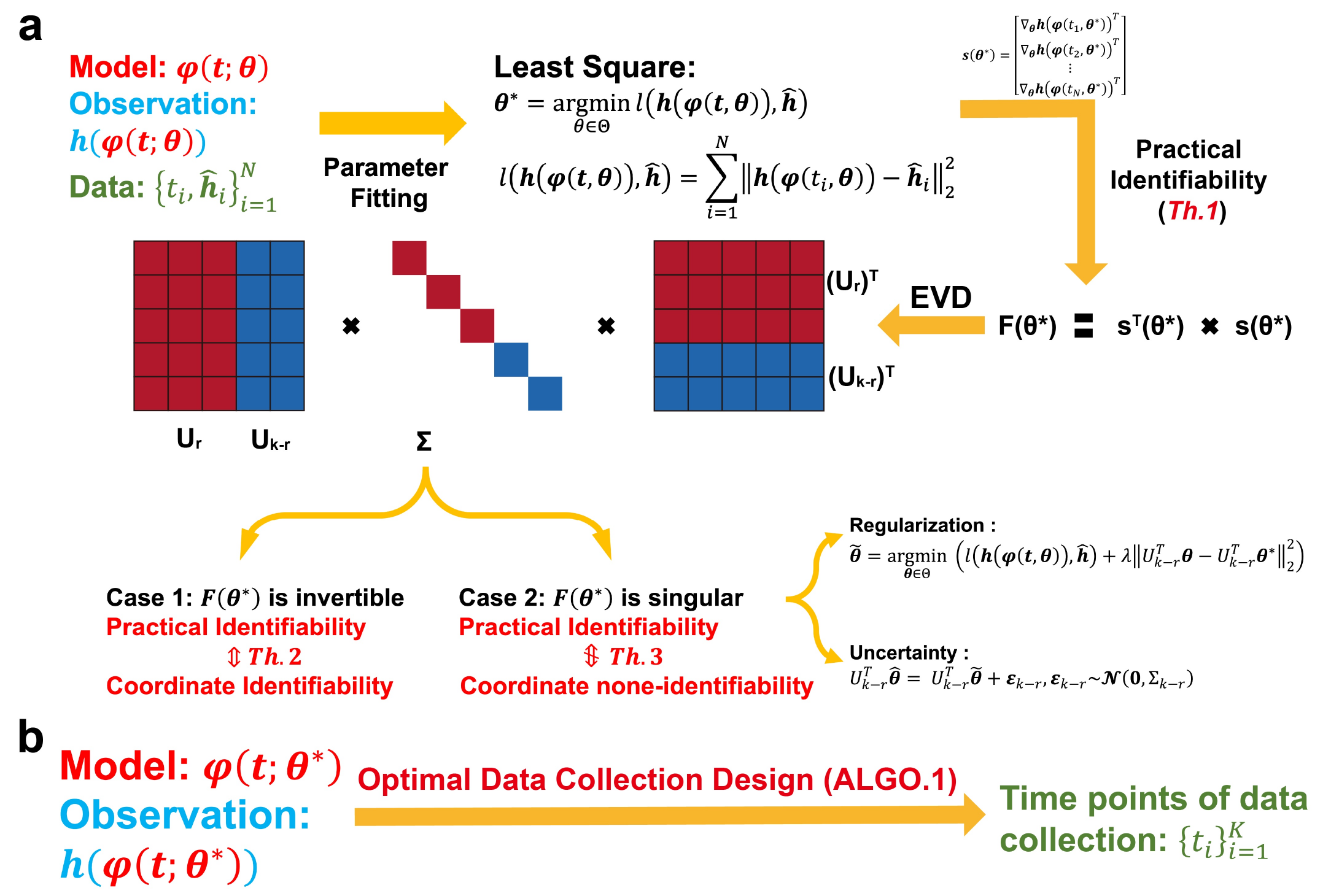}
\caption{\textbf{Illustration of the contributions presented in this study.} \textbf{(a)} A schematic representation of parameter practical identifiability analysis. Practical identifiability is determined by the eigenvalue matrix $\Sigma$, which is color-coded: red represents eigenvalues greater than zero, indicating practically identifiable, while blue represents eigenvalues equal to zero, signifying practically non-identifiable. In the eigenvector matrix U, the red portion corresponds to identifiable parameters, $\boldsymbol{U_r^T \theta}$, while the blue portion corresponds to non-identifiable parameters, $\boldsymbol{U_{k-r}^T \theta}$. \textbf{(b)} The optimization of data collection design informed by practical identifiability.}
\label{fig:1}
\end{figure}

To systematically perform practical identifiability analysis for model parameters, we propose a novel and rigorous mathematical definition of practical identifiability (Definition 1 in Materials and Methods).
This definition introduces the concept of practical identifiability from a data-fitting perspective, distinguishing it clearly  from the concept of structural identifiability. Practical identifiability analysis focuses on a general model $\boldsymbol{\varphi (t, \theta)}$, where $\boldsymbol{t}$ is the independent variable and $ \boldsymbol \theta \in \mathbb{R} ^k$ represents the parameter vector. The model may take any functional form or be the solution of differential equations, with observable variables $\boldsymbol{h(\varphi (t, \theta))}$ and experimental data collected at different time points $\{ t_i, \hat{\boldsymbol h}_i\}_{i=1}^N$. An initial parameter estimate, $\boldsymbol{\theta^*}$, is obtained via least-squares fitting. Next, we compute  the generalized parameter sensitivity matrix $\boldsymbol{s(\theta^*)}$ defined by Eq. \ref{eq:4} and the FIM as $F(\boldsymbol{\theta ^*}) = \boldsymbol{s^T(\theta^*)s(\theta^*)}$. Performing EVD on the FIM, Theorem 1 states that the parameter $\boldsymbol{\theta}$ is practically identifiable if and only if the FIM is invertible. Practical identifiability is therefore determined by the eigenvalue matrix $\Sigma$: eigenvalues greater than zero indicate that the corresponding parameter, $\boldsymbol{U_r^T \theta}$, are practically identifiable, while eigenvalues equal to zero mean that the corresponding parameters, $\boldsymbol{U_{k-r}^T \theta}$, is practically non-identifiable. This procedure is summarized in Figure \ref{fig:1}a.

We further explore the relationship between practical identifiability and coordinate identifiability (Definition 2 in Materials and Methods) through Theorems 2 and 3. Theorem 2 establishes their equivalence when the FIM is invertible, while Theorem 3 characterizes their differences in the singular FIM case.
To quantify identifiability capacity, we introduce the index $\| (I - AA^\dagger) \mathbf{s}_i \|_\infty$ where a lower value indicate a less identification for
parameter
$\theta_i \in \boldsymbol{\theta}$ (Details of A and $\boldsymbol{s_i}$ are in Theorem 3 in Materials and Methods). Moreover, when $F(\boldsymbol{\theta^*})$ is singular, some parameters are not practically identifiable. Thus, we propose a regularization method based on practical parameter identifiability to ensure that all parameters become practically identifiable during parameter fitting (Figure \ref{fig:1}a, details in ‘Parameter regularization’ section in Materials and Methods). Furthermore, for non-identifiable parameters, we develop a quantitative method to assess the uncertainty they introduce and evaluate their impact on model predictions (Figure \ref{fig:1}a, details in ‘Uncertainty Quantification’ section in Materials and Methods).

Building on the theorems and properties derived from analyzing the practical identifiability of model parameters (Figure \ref{fig:1}a), we propose a novel algorithm for designing experiments to ensure that the observed data renders all model parameters practically identifiable (Figure \ref{fig:1}b, Algorithm 1 in Materials and Methods). Using data or prior empirical information, initial model parameter $\boldsymbol{\theta^*}$ can be obtained as inputs to the algorithm. Algorithm 1 then generates a set of time points, representing the moments during the experiment when data measurements should be collected to ensure that all model parameters are practically identifiable.

\subsection*{Polynomial Fitting Benchmark Example }
To evaluate the accuracy of our proposed method, we apply it to a polynomial example as $h(t; \boldsymbol{\theta}) = \theta_1 + \theta_2 t^2 + \theta_3 \big[(t-1)(t-2)(t-3) + 2\big]
$ (more details can be found in Suppelmentary Materials) to compute practical identifiability and compare the results with the profile likelihood method \cite{raue2009structural,raue2013joining} which serves as a benchmark in practical identifiability analysis. 

\begin{figure}[ht]
\centering
\includegraphics[width=0.8\linewidth]{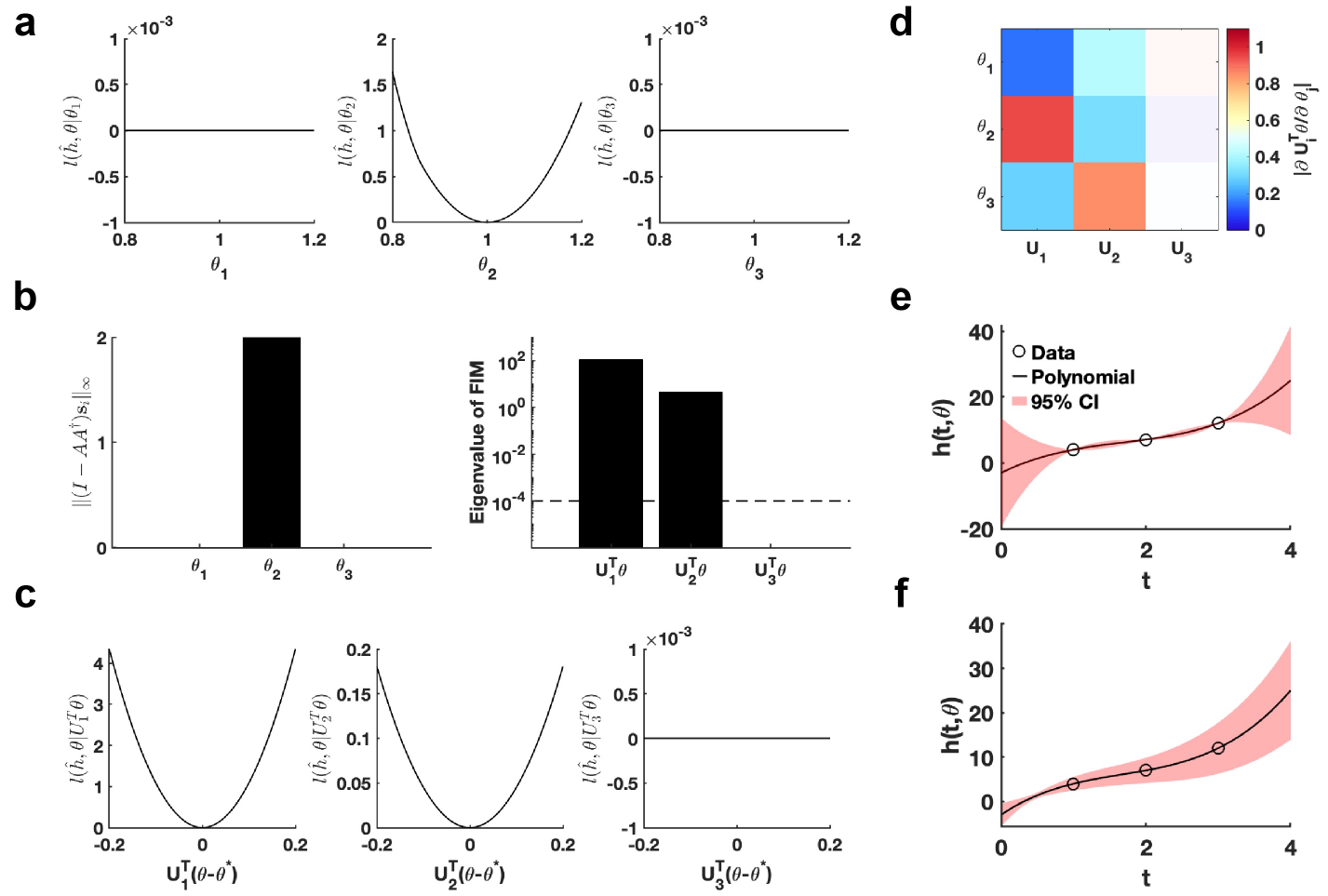}
\caption{\textbf{Validation method accuracy in polynomial fitting.} \textbf{(a)} Coordinate identifiability analysis at $\boldsymbol{\theta^*}=[1,1,1]^T$ using the profile likelihood. \textbf{(b)} Two metrics $\| (I - AA^\dagger) \mathbf{s}_i \|_\infty$ and eigenvalue of $F(\boldsymbol{\theta^*})$ for conducting practical identifiability analysis. The dashed line is the threshold $\varepsilon=10^{-4}$ of eigenvalue of $F(\boldsymbol{\theta^*})$. \textbf{(c)} Coordinate identifiability analysis to parameter $\boldsymbol{U^T\theta}$ using the profile likelihood. \textbf{(d)} Heatmap of the eigenvector matrix. The color bar represents the values of each eigenvector element. The shaded area indicates the eigenvectors corresponding to non-identifiable parameters. \textbf{(e)} Uncertainty quantification from the perturbation to non-identifiable parameters. \textbf{(f)} Uncertainty quantification from the perturbation to all parameters. Circles represent the synthetic data generated from the polynomial function. The solid line represents the polynomial function with the given parameter values $\boldsymbol{\theta^*}$. The red area represents the 95\% confidence interval under parameter perturbation.}
\label{fig:2}
\end{figure}

Using the given parameter $\boldsymbol{\theta^*}$, we utilize the profile likelihood method \cite{raue2013joining} to assess the identifiability of each parameter in the polynomial function, establishing a benchmark for comparison (Figure \ref{fig:2}a). Our proposed method computes two metrics, $\| (I - AA^\dagger) \mathbf{s}_i \|_\infty$ and eigenvalue of $F(\boldsymbol{\theta^*})$, to evaluate the practical identifiability of the polynomial function parameters. The result of $\| (I - AA^\dagger) \mathbf{s}_i \|_\infty$ demonstrates that only parameter $\theta_2$ is identifiable, whereas $\theta_1$ and $\theta_3$ are non-identifiable (Figure \ref{fig:2}b), consistent with the benchmark results (Figure \ref{fig:2}a). The eigenvalue of $F(\boldsymbol{\theta^*})$ further reveals the practical identifiability. Specifically, $\boldsymbol{U_1^T\theta}$ and $\boldsymbol{U_2^T\theta}$ are identifiable while the parameter $\boldsymbol{U_3^T\theta}$ is non-identifiable (Figure \ref{fig:2}b). To emphasize the discrepancy with the profile likelihood method, we perform a linear transformation on the parameters, namely, $\boldsymbol{U^T(\theta-\theta^*)}$. Then we conduct further practical identifiability analysis on the parameters using the profile likelihood method, which indicates that parameters $\boldsymbol{U_1^T\theta}$ and $\boldsymbol{U_2^T\theta}$ are identifiable, whereas parameter $\boldsymbol{U_3^T\theta}$ is non-identifiable (Figure \ref{fig:2}c). These results align perfectly with the parameter identifiability analysis shown in Figure \ref{fig:2}b but shows the sensitivity of the profile likelihood method. Leveraging this matrix $\boldsymbol{U}$, we incorporate a regularization term into the loss function to ensure that each parameter achieves practical identifiability. Subsequently, we perform the profile likelihood method to assess the identifiability of each parameter in the regularized loss function. The results confirm that all parameters become identifiable following the inclusion of the regularization term (Figure S1 in Supplementary Materials).

Finally, we introduce parameter perturbations and calculate the 95\% confidence interval for variations in the dependent variable. As shown in Figure \ref{fig:2}e, the confidence interval is nearly zero at the data points, indicating that the loss function remains unaffected by perturbations to the non-identifiable parameter only. Conversely, the result presented in Figure \ref{fig:2}f shows that the perturbations to all parameter influence the loss function at the data points, confirming that the loss function changes in response to perturbations to all parameters. This result highlights that our proposed uncertainty quantification method more precisely captures the prediction errors arising from parameter uncertainty. This accuracy is achieved because our method maintains the loss function's minimum under parameter perturbations.

\subsection*{Hill Functions and Neural Networks}
Next, we perform our proposed parameter practical identifiability analysis method to Hill functions and neural network functions, two widely used nonlinear models in systems biology. The primary objective is to determine whether our method can uncover the biological insights underlying these nonlinear functional classes.
\begin{figure}[ht]
\centering
\includegraphics[width=0.6\linewidth]{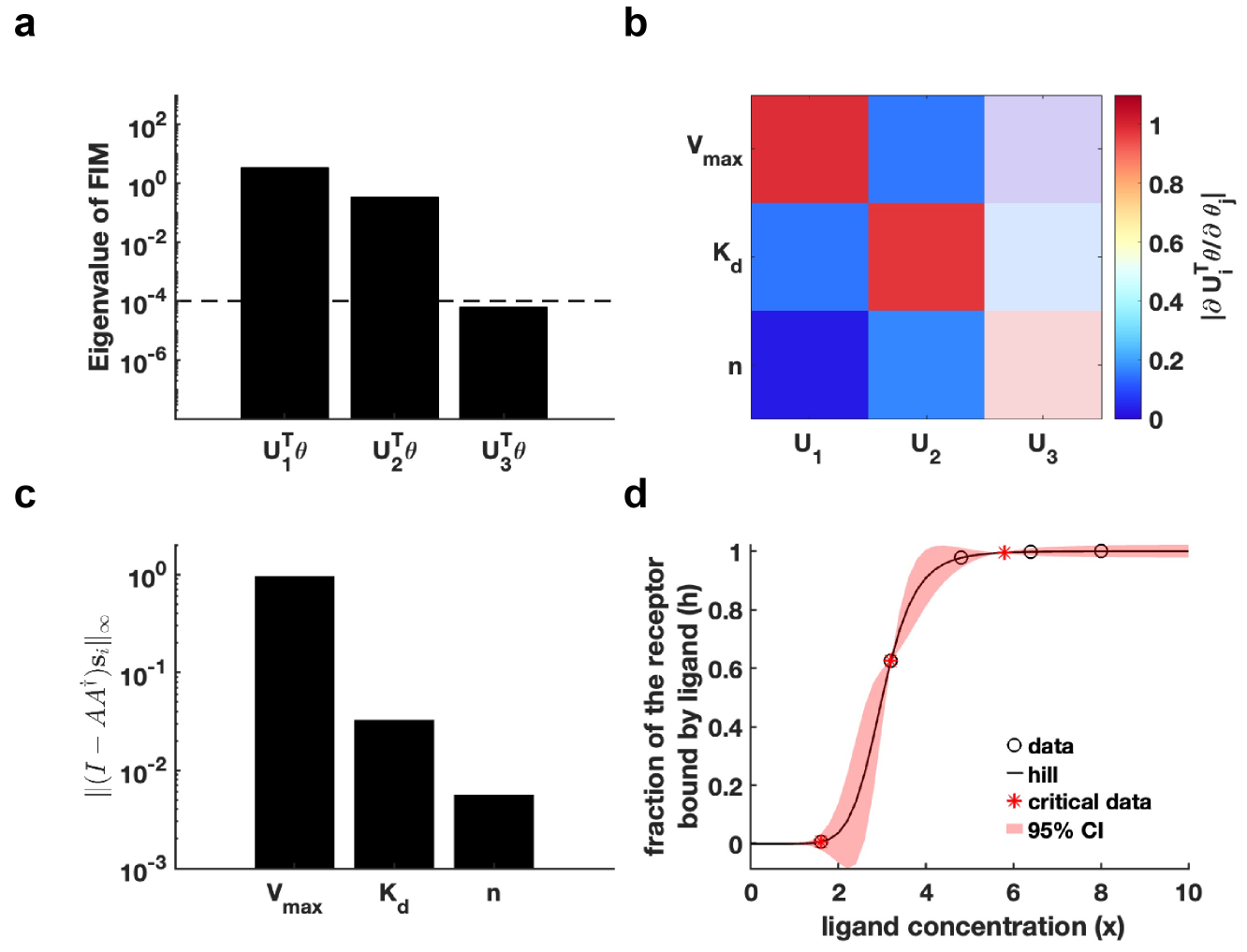}
\caption{\textbf{Practical identifiability analysis to Hill function.} \textbf{(a)} Eigenvalue of $F(\boldsymbol{\theta^*})$. The dashed line is the threshold $\varepsilon=10^{-4}$ of eigenvalue of $F(\boldsymbol{\theta^*})$. \textbf{(b)} Heatmap of the eigenvector matrix. The color bar represents the values of each eigenvector element. The shaded area indicates the eigenvectors corresponding to non-identifiable parameters. \textbf{(c)} Coordinate identifiability analysis to parameter $\boldsymbol{\theta}$ using the metric $\| (I - AA^\dagger) \mathbf{s}_i \|_\infty$. \textbf{(d)} Uncertainty quantification from the perturbation to non-identifiable parameters. Circles represent the synthetic data generated from the Hill function. The solid line represents the Hill function with the given parameter values $\boldsymbol{\theta^*}$. The red area represents the 95\% confidence interval under parameter perturbation. The star represents the critical data identified by algorithm 1 (Details in Materials and Methods).}
\label{fig:3}
\end{figure}

First, we generated a synthetic dataset using the predefined parameter \(\boldsymbol{\theta^*}\) for the Hill function  
\[ h(x; \boldsymbol{\theta}) = \frac{V_{\max} x^n}{x^n + K_d^n}, \]  
(Details in Supplementary Materials). We set an eigenvalue threshold \(\varepsilon=10^{-4}\) to classify eigenvectors with eigenvalues below this threshold as corresponding to non-identifiable parameters. The analysis in Figure \ref{fig:3}a reveals that the parameter \(\boldsymbol{U_3}^T \boldsymbol{\theta}\) is non-identifiable, whereas \(\boldsymbol{U_1}^T \boldsymbol{\theta}\) and \(\boldsymbol{U_2}^T \boldsymbol{\theta}\) are practically identifiable. Examination of the eigenvector matrix further confirms that parameters \(V_{\max}\) and \(K_d\) are identifiable, while parameter \(n\) is non-identifiable (Figure \ref{fig:3}b). Using the metric \(\| (I - AA^\dagger) \mathbf{s}_i \|_\infty\), we find that parameter \(V_{\max}\) exhibits the highest practical identifiability, followed by \(K_d\), with \(n\) showing the lowest identifiability (Figure \ref{fig:3}c). Furthermore, we employ the profile likelihood method as a benchmark for parameter identifiability analysis of the Hill function, yielding results fully consistent with our proposed method (Figure S2a in Supplementary Materials). Our practical identifiability analysis indicates that parameter \(n\) in the Hill function requires prior biological information for reliable estimation, as data fitting alone cannot precisely identify its value.

\begin{figure}[ht]
\centering
\includegraphics[width=0.8\linewidth]{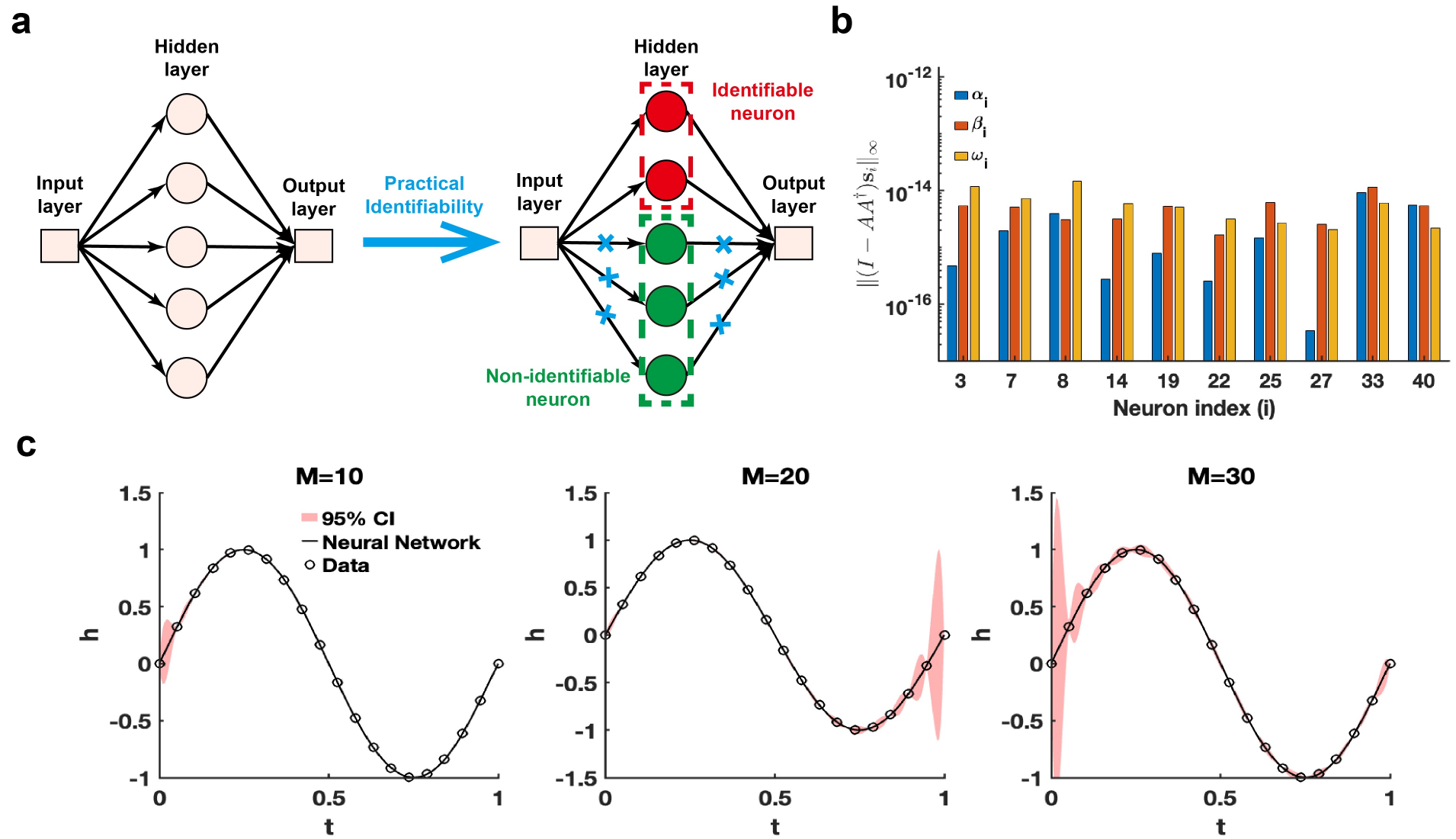}
\caption{\textbf{Practical identifiability analysis of neural network with one hidden layer.} \textbf{(a)} Schematic of parameter practical identifiability applied in neural networks. \textbf{(b)} Identifiable neurons recognized by the metric $\| (I - AA^\dagger) \mathbf{s}_i \|_\infty$ when the activation function set to the ReLu function and the number of neurons is assigned as 40. \textbf{(c)} Uncertainty quantification was performed by introducing perturbations to non-identifiable parameters across different numbers of neurons (M), with the activation function set to the tanh function. Circles represent the synthetic data generated from the sine function $sin(2\pi t),t\in[0,1]$. The solid line represents the neural network with the given parameter values $\boldsymbol{\theta^*}$. The red area represents the 95\% confidence interval under non-identifiable parameter perturbation.}
\label{fig:4}
\end{figure}

Finally, we introduce parameter perturbations and compute the 95\% confidence interval of the dependent variable. The results in Figure \ref{fig:3}d show that perturbations to the non-identifiable parameter \(n\) primarily affect the region near the Hill function's inflection point. Compared to the confidence interval from perturbing all parameters (Figure S2b in Supplementary Materials), Figure \ref{fig:3}d more accurately reflects the actual data-fitting process of the Hill function. Moreover, we utilize Algorithm 1 to determine the critical ligand concentration points of the Hill function, identifying these points that render all parameters practically identifiable (Figure \ref{fig:3}d).

The single hidden-layer neural network is constructed to fit the sine function \(\sin(2\pi t), t \in [0,1]\), leveraging practical parameter identifiability analysis to identify neurons with practically identifiable parameters. For neurons deemed non-identifiable, regularization terms are introduced to fix their parameters during training, enabling the model to focus exclusively on training parameters of identifiable neurons (Figure \ref{fig:4}a). This approach shows promise for accelerating training and improving prediction accuracy. When the activation function is ReLU and the number of neurons is set to 40, we use the metric \(\| (I - AA^\dagger) \mathbf{s}_i \|_\infty\) to recognize identifiable neurons (Figure \ref{fig:4}b). The remaining neurons are classified as non-identifiable because their corresponding metric values are zero. When switching the activation function to tanh, the metrics \(\| (I - AA^\dagger) \mathbf{s}_i \|_\infty\) for all neurons are positive (Figure S3a in Supplementary Materials). Furthermore, 95\% confidence intervals are computed for varying neuron counts, revealing that uncertainty increases with more neurons (Figure \ref{fig:4}c). The eigenvalue distribution across varying neuron numbers (Figure S3b in Supplementary Materials) shows that the proportion of eigenvalues exceeding the threshold decreases as neuron count increases. Whether the 95\% confidence interval widens as $t$ approaches 0 or 1 depends on the random initialization of the neural network parameters (Figure S8 in Supplementary Materials). The findings presented indicate that an excessive number of neurons in a single-layer neural network heightens parameter-induced uncertainty, potentially slowing down the training process and increasing the risk of the Runge phenomenon.

\section*{Various Biological Systems with differential equations.}
\subsection*{LV model}
\begin{figure}[ht]
\centering
\includegraphics[width=0.8\linewidth]{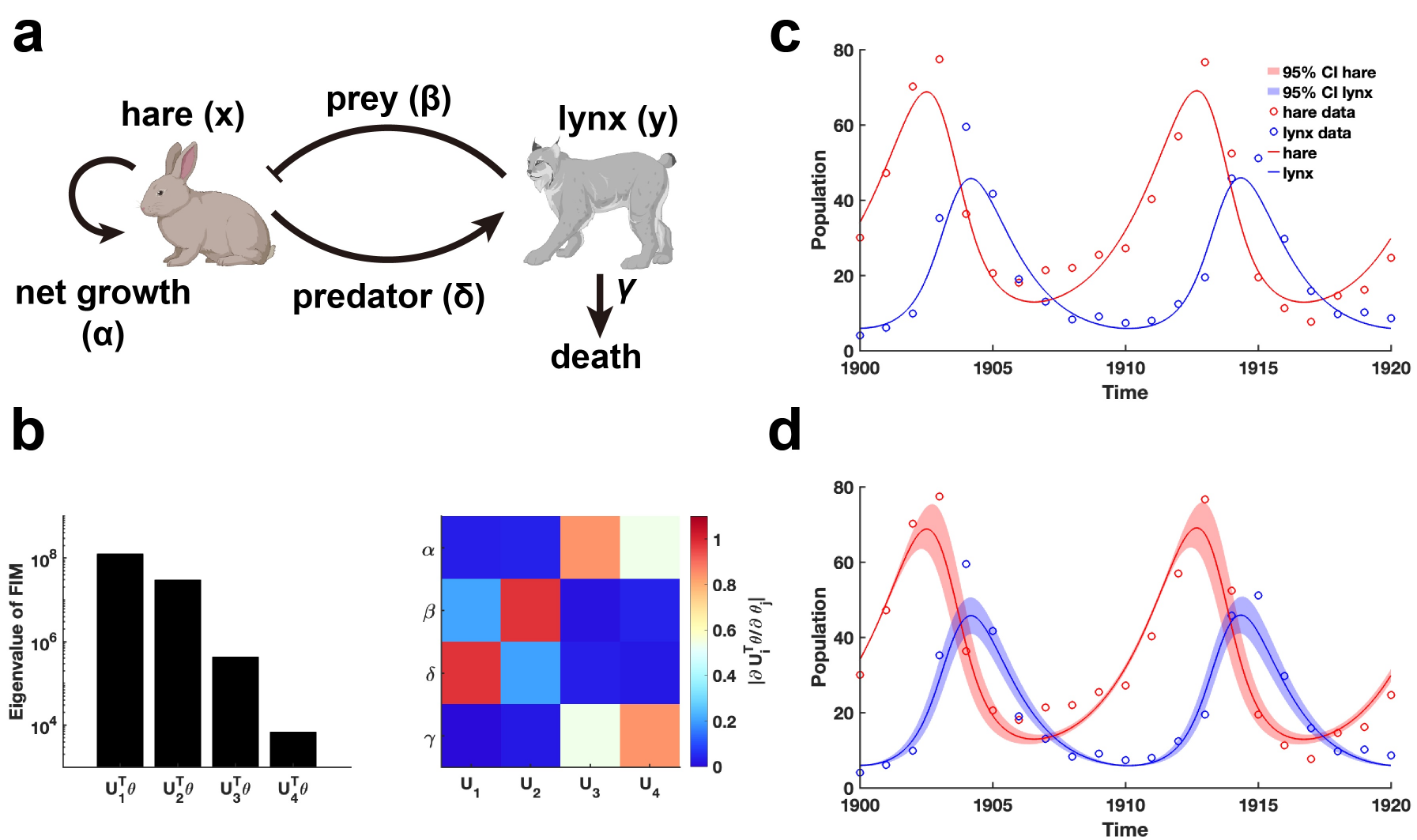}
\caption{\textbf{Practical identifiability analysis of LV model.} \textbf{(a)} Schematic of LV model. \textbf{(b)} Eigenvalue of $F(\boldsymbol{\theta^*})$ and heatmap of the eigenvector matrix. The color bar represents the values of each eigenvector element. The parameter $\boldsymbol{\theta^*}$ values are provided in the "Parameter Values" section of the Supplementary Materials. \textbf{(c)} Uncertainty quantification is performed by introducing perturbations to non-identifiable parameters. \textbf{(d)} Uncertainty quantification from the perturbation to all parameters. Circles represent the real data of hare and lynx obtained from published literature \cite{howard2009modeling}. The solid line represents the LV model with the given parameter values $\boldsymbol{\theta^*}$. The red area represents the 95\% confidence interval under parameter perturbation.}
\label{fig:5}
\end{figure}

We begin by examining the classic predator-prey relationship within ecological network models using the LV model \cite{murray2002mathematical} (Figure \ref{fig:5}a). Public data on hare and lynx populations \cite{howard2009modeling} are utilized for parameter estimation through data fitting. Using the obtained parameters $\boldsymbol{\theta^*}$, we calculate the FIM $F(\boldsymbol{\theta^*})$ and conduct EDV to derive the eigenvalues and their corresponding eigenvectors (Figure \ref{fig:5}b). Our analysis reveals that parameters $(\beta,\delta)$ associated with the predator-prey interaction exhibited the highest eigenvalues, followed by the intrinsic growth and death rates of the species (Figure \ref{fig:5}b). This finding indicates that the periodic fluctuations observed in the hare and lynx populations are predominantly driven by the predator-prey interaction parameters, emphasizing their role in inducing periodic dynamics. Moreover, the invertibility of the FIM confirm that the parameters are uniquely identifiable without uncertainty (Figure\ref{fig:5}c and Theorem 1). Although the confidence intervals derived from perturbing all parameters exhibit periodic variations (Figure \ref{fig:5}d), the perturbed parameters failed to preserve the loss function in data fitting.

\subsection*{Michaelis–Menten system} 
\begin{figure}[ht]
\centering
\includegraphics[width=0.85\linewidth]{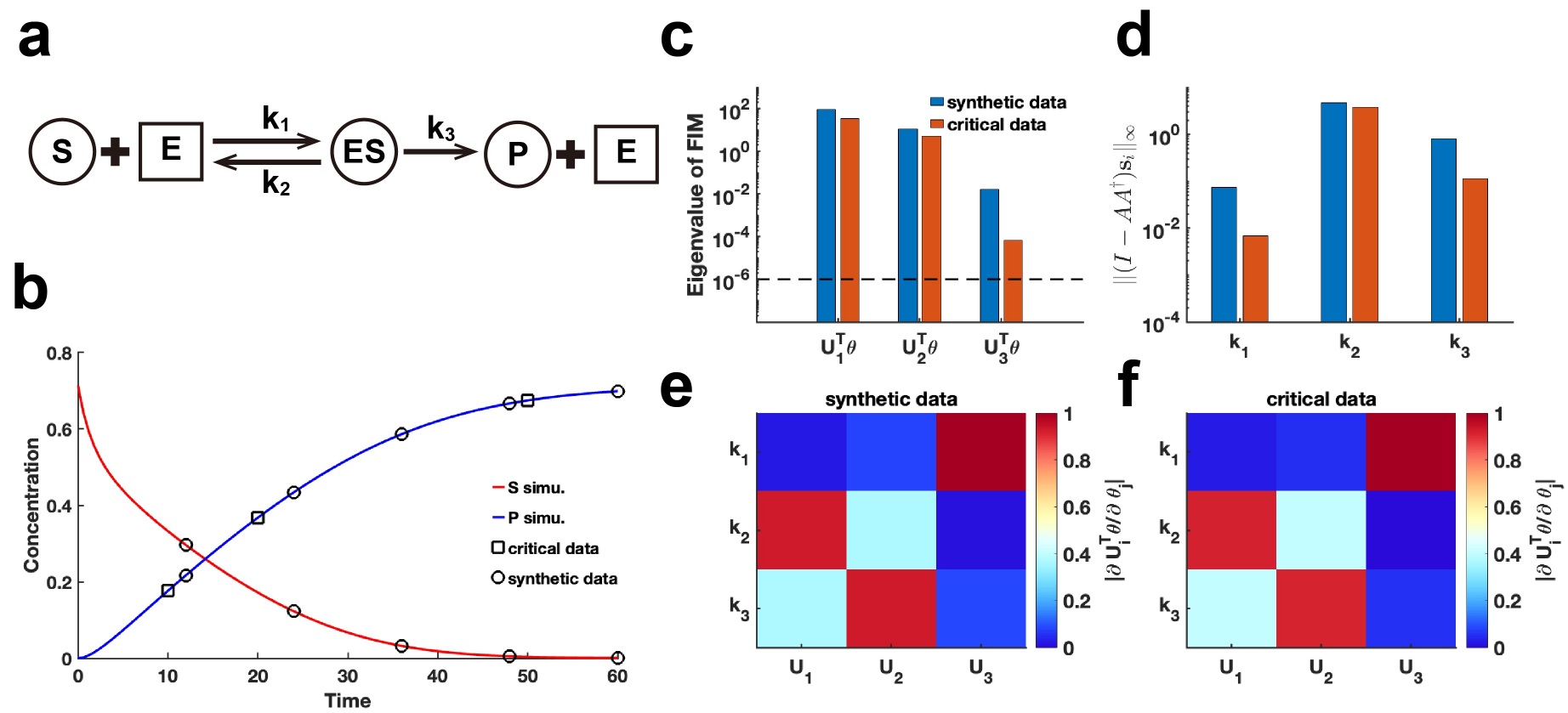}
\caption{\textbf{Practical identifiability analysis of Michaelis–Menten system.} \textbf{(a)} Schematic of Michaelis–Menten system. S, E, ES, and P represent substrate, enzymes, complex of substrate and enzymes, and product, respectively. \textbf{(b)} Time course of substrate and product. Circles represent the synthetic data generated by the given parameter values $\boldsymbol{\theta^*}$ (44). The solid line represents the Michaelis–Menten model with the given parameter values $\boldsymbol{\theta^*}$. Squares represent the critical data identified by Algorithm 1. \textbf{(c)} Eigenvalue of $F(\boldsymbol{\theta^*})$ using the synthetic data and critical data respectively. The dash line is the threshold $\varepsilon=10^{-6}$. \textbf{(d)} Coordinate identifiability analysis to parameter $\boldsymbol{\theta^*}$ using the metric $\| (I - AA^\dagger) \mathbf{s}_i \|_\infty$ for the synthetic data and critical data. \textbf{(e)} Heatmap of the eigenvector matrix using synthetic data. \textbf{(f)} Heatmap of the eigenvector matrix using synthetic data. The color bar represents the values of each eigenvector element. The parameter $\boldsymbol{\theta^*}$ values are provided in the "Parameter Values" section of the Supplementary Materials.}
\label{fig:6}
\end{figure}

We extend our method to assess the practical identifiability of parameters in the classic enzyme-catalyzed reaction model, the Michaelis–Menten system \cite{lei2021systems} (Figure\ref{fig:6}a). Using parameters $\boldsymbol{\theta^*}$ obtained from the literature \cite{higham2008modeling}, we generate synthetic data with the observable variable set as the substrate and product concentration (Figure \ref{fig:6}b) (case (1) of Michaelis–Menten system in Supplementary Materials). Additionally, we alter the observable variable to the product concentration (case (2) of Michaelis–Menten system in Supplementary Materials) and perform Algorithm 1 to identify critical data points.
By comparing the eigenvalue distributions of the FIM $F(\boldsymbol{\theta^*})$ generated using synthetic data and critical data, we consistently observe that parameter $k_1$ exhibited the lowest identifiability (Figures \ref{fig:6}c, e, f). Using our proposed metric $\| (I - AA^\dagger) \mathbf{s}_i \|_\infty$ to analyze coordinate identifiability, we confirm that $k_1$ has the lowest identifiability among the parameters. This finding highlights the difficulty in accurately capturing the chemical reaction constant associated with the binding of the enzyme to the substrate, regardless of whether substrate or product data are utilized. Moreover, the eigenvector matrices derived from both data types are identical (Figures \ref{fig:6}e-f), indicating that Algorithm 1 effectively guides experimental design for optimizing data measurement in the Michaelis–Menten system. We examine our framework using the synthetic data with additive Gaussian noise for Michaelis-Menten system. The results present that both practical and coordinate identifiability almost remain unchanged in the presence of this noise, demonstrating the robustness of our framework with respect to parameter identifiability (Figure S7 in Supplementary Materials).

\subsection*{SEIR model} 
\begin{figure}[ht]
\centering
\includegraphics[width=0.83\linewidth]{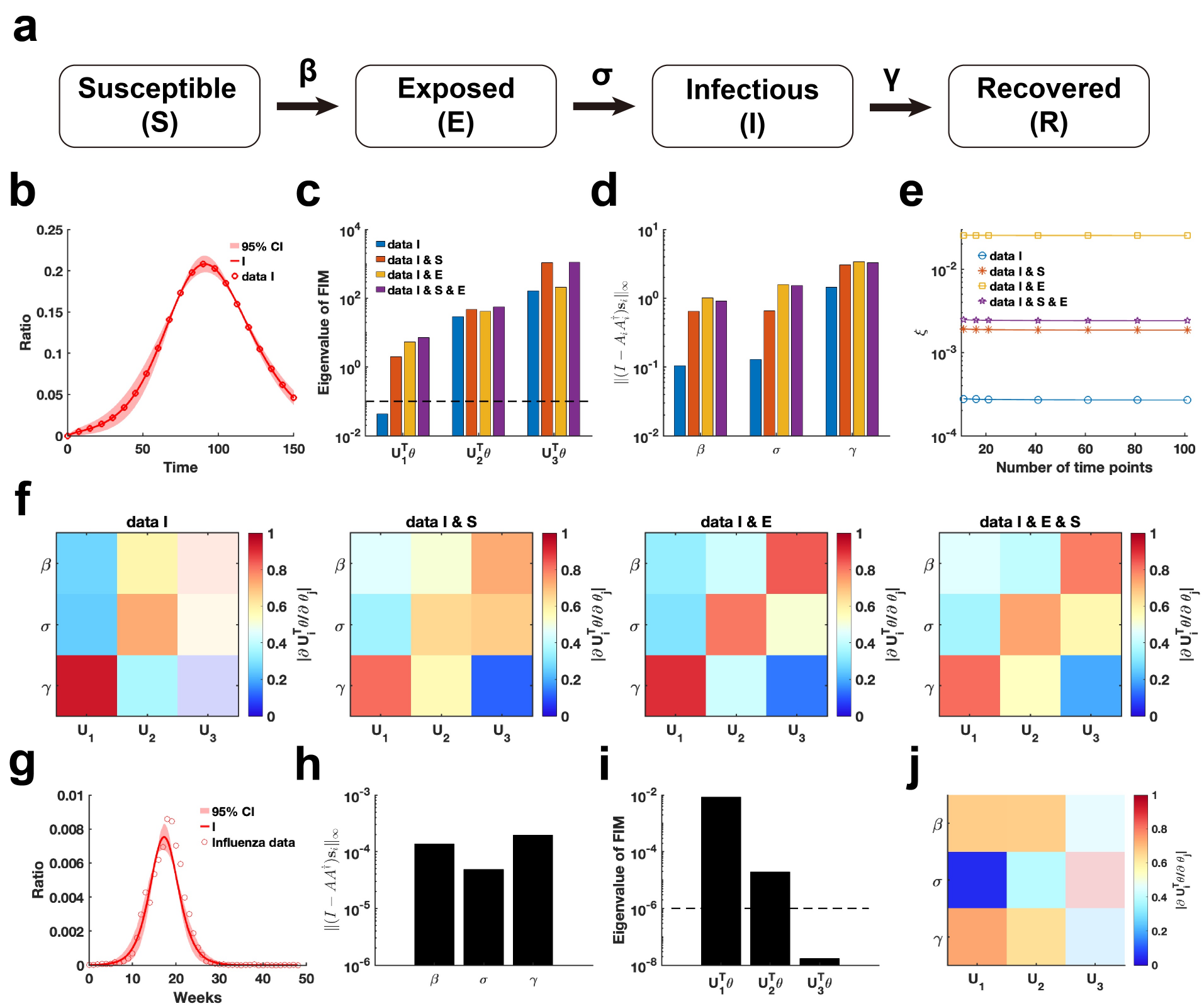}
\caption{\textbf{Practical identifiability analysis of SEIR model.} \textbf{(a)} Schematic of SEIR model. \textbf{(b)} Uncertainty quantification is performed by introducing perturbations to non-identifiable parameters. Circles represent the synthetic data of Infected ratio when the parameter values are given. The solid line represents the infected ratio of SEIR model with the given parameter values. The red area represents the 95\% confidence interval under parameter perturbation. \textbf{(c)} Eigenvalue of FIM in the four cases of observable variables (Details in Supplementary Materials). The dash line is the threshold $\varepsilon=10^{-6}$. \textbf{(d)} Coordinate identifiability analysis to parameter using the metric $\| (I - AA^\dagger) \mathbf{s}_i \|_\infty$ for four cases of observable variables. \textbf{(e)} Contribution of different data types to parameter practical identifiability using the metric $\xi$ across multiple time points. \textbf{(f)} Heatmap of the eigenvector matrix in the four cases of observable variables. \textbf{(g)} Uncertainty quantification is performed by introducing perturbations to non-identifiable parameters. Circles represent the influenza data of Infected ratio obtained from the CDC website. The solid line represents the infected ratio of SEIR model with the given parameter values. The red area represents the 95\% confidence interval under parameter perturbation. \textbf{(h)} Coordinate identifiability analysis to parameter using the metric $\| (I - AA^\dagger) \mathbf{s}_i \|_\infty$ for the influenza data. \textbf{(i)} Eigenvalue of FIM using the influenza data. The dash line is the threshold $\varepsilon=10^{-6}$. \textbf{(j)} Heatmap of the eigenvector matrix using the influenza data. The color bar represents the values of each eigenvector element. The parameter values are provided in the "Parameter Values" section of the Supplementary Materials.}
\label{fig:7}
\end{figure}

We employ our proposed parameter practical identifiability method to investigate the SEIR infectious disease model \cite{hunter2022understanding}, a system distinguished by its greater number of state variables compared to parameters (Figure \ref{fig:7}a). First, we utilized synthetic data to evaluate the practical identifiability of the model parameters. With the observable variable designated as $h(t,\boldsymbol{\theta})=I(t,\boldsymbol{\theta})$, synthetic data are generated using a specific parameter $\boldsymbol{\theta^*}$ (See in the "Parameter Values" section of the Supplementary Materials). Parameter uncertainty analysis based on the synthetic data indicated that the uncertainty in infected patient data is notably higher during the early stages of the outbreak (Figure \ref{fig:7}b). In contrast, uncertainty analysis conducted by perturbing all parameters demonstrates nearly zero uncertainty in the infected patient data during the early stages of the outbreak (Figure S4 in Supplementary Materials).

Next, we analyze the eigenvalue distributions of the FIM (Figure \ref{fig:7}c) and evaluate the coordinate identifiability of parameters across four different scenarios of observable variables using the metric $\| (I - AA^\dagger) \mathbf{s}_i \|_\infty$ (Figure \ref{fig:7}d). Our findings demonstrate that increasing the number of observable variables ensures that all parameters become practically identifiable and significantly enhances the identifiability of each parameter in the model. When the observable variable is set to $h(t,\boldsymbol{\theta})=[E(t,\boldsymbol{\theta}),I(t,\boldsymbol{\theta})]$, the data provides the highest contribution to parameter identifiability within the model (Figure \ref{fig:7}d). This suggests that, in the SEIR model, focusing on monitoring exposed and infected individuals is sufficient for accurately predicting the later stages of an epidemic. Additionally, a comparison of the eigenvector matrices showed that, with $h(t,\boldsymbol{\theta})=[E(t,\boldsymbol{\theta}),I(t,\boldsymbol{\theta})]$ as the observable variable, the weight of each eigenvector is concentrated on a single parameter (Figure \ref{fig:7}f). This result underscores the importance of monitoring exposed and infected individuals, as it maximizes the identifiability of individual parameters within the SEIR model.

Finally, we estimated the parameters of the SEIR model using influenza A data from the 2004-2005 season, obtained from the CDC website (Details in Data Availability), and analyzed the practical identifiability of the estimated parameters. Uncertainty analysis reveals that the model predictions exhibit the highest levels of uncertainty during the initial stages and at the peak of the influenza outbreak (Figure \ref{fig:7}g). Using metric $\| (I - AA^\dagger) \mathbf{s}_i \|_\infty$, it shows that the identifiability of transmission rate $(\beta)$ and the recovery rate $(\gamma)$ is nearly identical, while the incubation rate $(\sigma)$ exhibits the lowest identifiability (Figure \ref{fig:7}h). The eigenvalue distribution of the FIM and the corresponding eigenvector matrix further confirm the low identifiability of the incubation rate $(\sigma)$ (Figures \ref{fig:7}i-j). These findings underscore the critical importance of monitoring exposed individuals to enhance the predictive accuracy of the SEIR model.

\subsection*{Cascade model of Alzheimer’s Disease}
\begin{figure}[ht]
\centering
\includegraphics[width=1.0\linewidth]{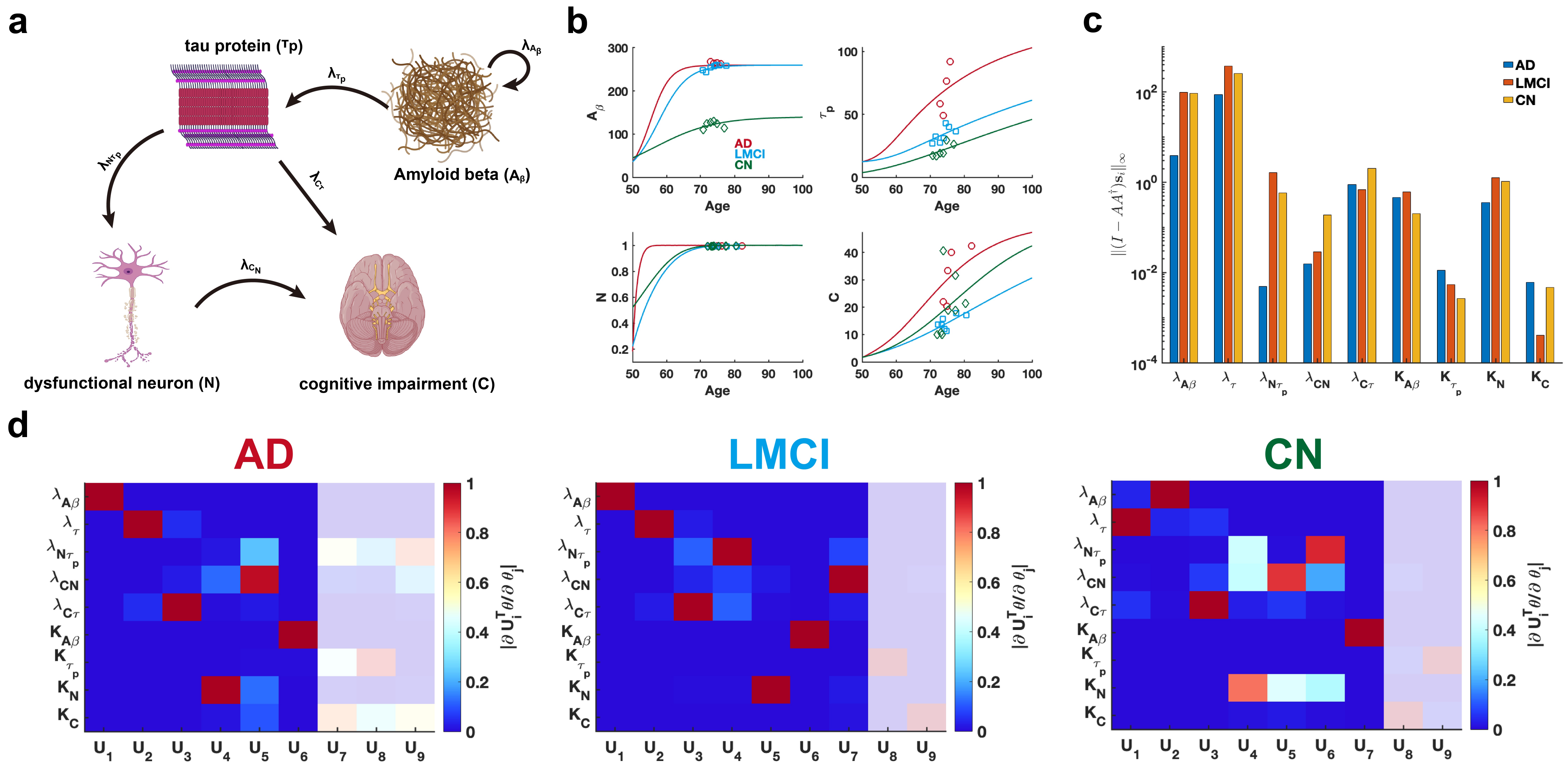}
\caption{\textbf{Practical identifiability analysis to cascade model of Alzheimer’s Disease.} \textbf{(a)} Schematic of cascade model of Alzheimer’s Disease.  \textbf{(b)} Time course of four biomarkers. Circles, squares, and diamonds represent the real data of AD, LMCI and CN patients, respectively. The red, orange, and green solid line represents the time course of four biomarkers with the given parameter values. \textbf{(c)} Coordinate identifiability analysis to parameter using the metric $\| (I-AA^\dagger) \mathbf{s}_i \|_\infty$. \textbf{(d)} Heatmap of the eigenvector matrix using the real data of AD, LMCI and CN patients. The shaded area indicates the eigenvectors corresponding to non-identifiable parameters. The color bar represents the values of each eigenvector element. All parameter values are provided in the "Parameter Values" section of the Supplementary Materials.}
\label{fig:8}
\end{figure}

We conduct a practical parameter identifiability analysis on biomarker cascade model of Alzheimer’s Disease (AD) \cite{hao2022optimal}, incorporating data from three clinical groups: cognitively normal (CN), late mild cognitive impairment (LMCI), and AD (Figure \ref{fig:8}a) from Alzheimer's Disease Neuroimaging Initiative(ADNI) dataset. The primary goal is to leverage practical identifiability analysis to identify variations in model parameters across these groups, thereby uncovering critical biological processes that distinguish the clinical conditions.

Using time-series data of four biomarkers from AD, LMCI, and CN patients, we reconstruct disease progression trajectories for different patient groups (Figure \ref{fig:8}b). Using metric  $\| (I - AA^\dagger) \mathbf{s}_i \|_\infty$, we assess the identifiability of each parameter across the three clinical symptom groups and identify two key parameters, growth rate of N $(\lambda_{N\tau_p})$ and carrying capacity of C $(K_C)$, that significantly distinguish these groups (Figure \ref{fig:8}c). Our analysis reveals that parameter $(\lambda_{N\tau_p})$ demonstrates substantially lower identifiability in AD patients compared to CN and LMCI groups, whereas parameter $K_C$ exhibits markedly reduced identifiability in LMCI patients relative to the others (Figure \ref{fig:8}c). These findings suggest that evaluating the identifiability of parameters $(\lambda_{N\tau_p})$ and $K_C$ within the cascade model provides a robust means of distinguishing between CN, LMCI, and AD patients.

 By establishing a threshold for the eigenvalues (Figures S5a-c in Supplementary Materials), we observe that AD patients exhibit a greater number of non-identifiable parameters compared to LMCI and CN groups. These findings imply that, given comparable data types and quantities, patients with a higher proportion of non-identifiable parameters identified through FIM analysis are more likely to be diagnosed with AD.

To investigate the influence of data from different age phases on practical identifiability, 
we perform coordinate identifiability analysis using synthetic data stratified by age, 
with model parameters held fixed. The results demonstrate that parameters such as 
$\lambda_{A_\beta}$, $\lambda_{N_{\tau_p}}$, $\lambda_{\mathrm{CN}}$, 
$\lambda_{C_\tau}$, and $K_C$ are highly identifiable in the early-age phase, 
whereas parameters such as $\lambda_\tau$, $K_{\tau_p}$, and $K_{A_\beta}$ 
exhibit higher identifiability in the later-age phase (Figure~S9 in Supplementary Materials).

\subsection*{PDE model of cancer-immune interactions}
\begin{figure}[ht]
\centering
\includegraphics[width=1.0\linewidth]{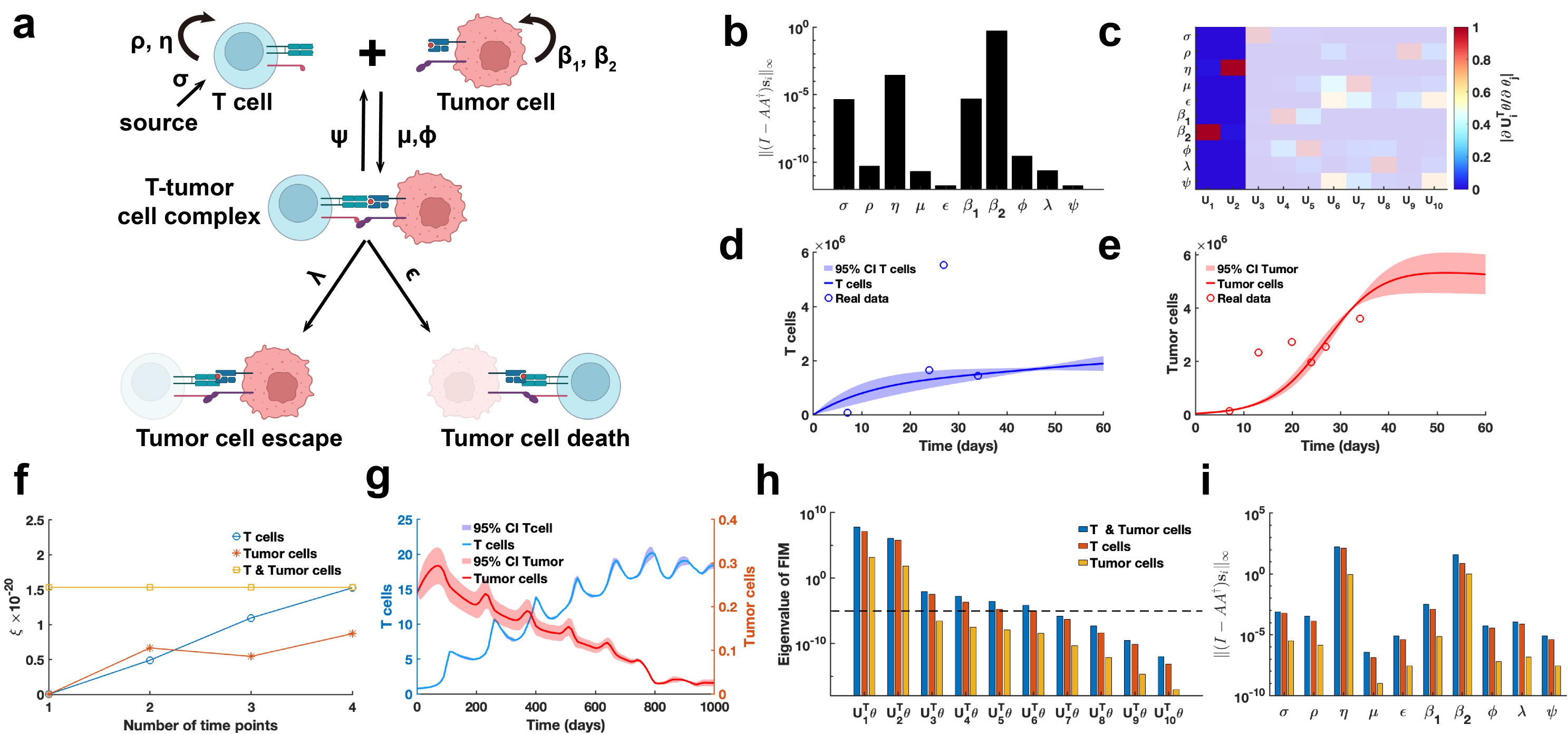}
\caption{\textbf{Practical identifiability analysis to PDE model of cancer-immune interactions.} \textbf{(a)} Schematic of PDE model of cancer-immune interactions. \textbf{(b)} Coordinate identifiability analysis to parameter using the metric $\| (I - AA^\dagger) \mathbf{s}_i \|_\infty$ based on glioblastoma data. \textbf{(c)} Heatmap of the eigenvector matrix using the glioblastoma data. The shaded area indicates the eigenvectors corresponding to non-identifiable parameters. The color bar represents the values of each eigenvector element. \textbf{(d)} Uncertainty quantification of T cells is performed by introducing perturbations to non-identifiable parameters. Circles represent the experimental data of T cells. The solid line represents the time course of T cells with the given parameter values. The red area represents the 95\% confidence interval under parameter perturbation. \textbf{(e)}. Uncertainty quantification of tumor cells is performed by introducing perturbations to non-identifiable parameters. Circles represent the experimental data of tumor cells. The solid line represents the time course of tumor cells with the given parameter values. The red area represents the 95\% confidence interval under parameter perturbation. \textbf{(f)}. Contribution of different data types to parameter practical identifiability using the metric $\xi$ across multiple time points. \textbf{(g)}. Uncertainty quantification of both T and tumor cells from the non-identifiable parameters. \textbf{(h)}. Eigenvalue of FIM in the three cases of observable variables (Details in Supplementary Materials). The dash line is the threshold $\varepsilon=10^{-5}$. \textbf{(i)}. Coordinate identifiability analysis to parameter using the metric $\| (I - AA^\dagger) \mathbf{s}_i \|_\infty$ for three cases of observable variables.}
\label{fig:9}
\end{figure}

As the final example, our proposed parameter identifiability method is employed to investigate the classic cancer-immune interaction PDE model (Figure \ref{fig:9}a) \cite{matzavinos2004mathematical}. In contrast to above analyses of biological system models, this model accounts for stochastic cell movement and intricate interaction mechanisms (Figure \ref{fig:9}a), thereby increasing the complexity of the parameter practical identifiability analysis. Our aim is to leverage practical identifiability analysis to uncover critical biological processes of cancer-immune interactions embedded in the model and to determine the key observable variables.

Using public glioblastoma data \cite{anderson2024global}, which included multiple time-point measurements of T cells and tumor cells, we estimated the parameters of the cancer-immune interaction PDE model, except for the tumor cell random movement parameter $\omega$, which was determined based on prior information. $h(t, \boldsymbol{\theta}) = 
[\int_0^1 E(t, x; \boldsymbol{\theta}) \, dx, 
\int_0^1 T(t, x; \boldsymbol{\theta}) \, dx] 
$ is observable variable for the glioblastoma data. Using metric $\| (I - AA^\dagger) \mathbf{s}_i \|_\infty$, we observe that parameters $(\sigma,\eta,\beta_1,\beta_2)$ with high identifiability are predominantly associated with the biological processes of T cell and tumor cell proliferation and apoptosis (Figure \ref{fig:9}b). In contrast, parameters $(\mu,\epsilon,\phi,\lambda,\psi)$ linked to T cell-tumor cell interactions exhibit low identifiability. Based on the identifiability threshold (Figure S6a in Supplementary Materials), we identify that the most identifiable parameters are those related to T cell and tumor cell proliferation (Figure \ref{fig:9}c). Uncertainty quantification for T cells presents the high levels of uncertainty in their counts during the early stages of the process (Figure \ref{fig:9}d). Conversely, for tumor cells, the high uncertainty is observed in the later stages, where their counts stabilized at a steady-state level (Figure \ref{fig:9}e).

To investigate the influence of spatial cell movement on parameter practical identifiability, we generate synthetic data with spatial information based on predefined parameters. In this context, three scenarios of observable variables are analyzed (Details in Supplementary Materials). Using metric $\xi$, we evaluate the contributions of these variables to the identifiability of model parameters and find that T cell data provides greater contributions to parameter identifiability compared to tumor cell data (Figure\ref{fig:9}f). Uncertainty quantification using the synthetic data reveals that tumor cell counts exhibit greater uncertainty during the early stages (Figure \ref{fig:9}g). Analysis of the eigenvalue distribution of the FIM (Figure \ref{fig:9}h) and its corresponding eigenvector matrix (Figure S6b) shows that T cell data allows more parameters in the model to be practically identifiable compared to tumor cell data. Metric $\| (I - AA^\dagger) \mathbf{s}_i \|_\infty$ further demonstrates that T cell data renders parameters $(\mu,\epsilon,\phi,\lambda,\psi)$ related to T cell-tumor cell interactions practically identifiable, whereas tumor cell data primarily identifies parameters $(\sigma,\eta,\beta_1,\beta_2)$ associated with the proliferation of T cells and tumor cells (Figure \ref{fig:9}i). These results underscore the importance of prioritizing the collection of T cell data in practical experiments to improve the model’s capacity for accurately predicting cancer progression.

\section*{Discussion}
Practical identifiability is a fundamental aspect of mathematical modeling in biological systems, as it directly influences the reliability and robustness of model predictions. While practical identifiability has attracted substantial attention in the modeling community, many existing approaches rely on heuristic criteria, numerical approximations, or are associated with high computational cost, often resulting in potentially misleading conclusions regarding parameter identifiability and the reliability of model predictions.
In this paper, we propose a novel framework for practical identifiability analysis that integrates the concept of coordinate identifiability. Additionally, we introduce an optimal data collection algorithm that utilizes practical identifiability to guide experimental design, thereby improving the efficiency and precision of data acquisition.

We introduce a rigorous mathematical definition for practical identifiability (Definition 1 in Materials and Methods). While the invertibility of the FIM has often been used as a criterion for practical identifiability \cite{gallo2022lack,miao2011identifiability,komorowski2011sensitivity}, its theoretical foundation has remained unproven. Building on our proposed definition, we formally establish the relationship between parameter practical identifiability and the invertibility of the FIM (Theorem 1 in Materials and Methods). Additionally, we elucidate the relationship between practical identifiability and structural identifiability (Theorem 4 in Materials and Methods), which reveals that if the parameter $\boldsymbol{\theta}$ is structurally identifiable, we can discover the time series $\{t_{i_j}\}_{j=1}^M$ to make $\boldsymbol{\theta}$ practically identifiable. Consequently, structural identifiability analysis can be regarded as a limiting case of practical identifiability analysis when the dataset becomes infinitely large, such as in model-generated synthetic data. This insight suggests that structural identifiability can be effectively assessed by conducting practical identifiability analysis on sufficiently rich synthetic datasets.

Coordinate identifiability has received considerable attention in the analysis of dynamic models within systems biology. Traditionally, the profile likelihood method has been used to evaluate the identifiability of individual parameters \cite{wieland2021structural,raue2009structural,raue2013joining}. However, this method becomes computationally infeasible for high-dimensional models, such as the cell cycle signaling pathway model with 48 parameters \cite{brown2003statistical}, posing challenges for accurately assessing parameter identifiability. First, we establish that practical identifiability and coordinate identifiability are equivalent when the FIM is invertible (Theorem 2 in Materials and Methods). Second, for cases where the FIM is singular, we introduce a novel metric $\| (I - AA^\dagger) \mathbf{s}_i \|_\infty$ to evaluate the identifiability of individual parameters. We further demonstrate that this metric acts as a linear approximation to the profile likelihood method (Theorem 3 in Materials and Methods). Compared to the profile likelihood approach, our proposed method significantly reduces computational cost while offering a more precise analysis of parameter identifiability.

\begin{table}[ht]
  \centering
  \caption{ \bf Summary of practical identifiability analysis on various biological models.}
  \includegraphics[width=1\textwidth]{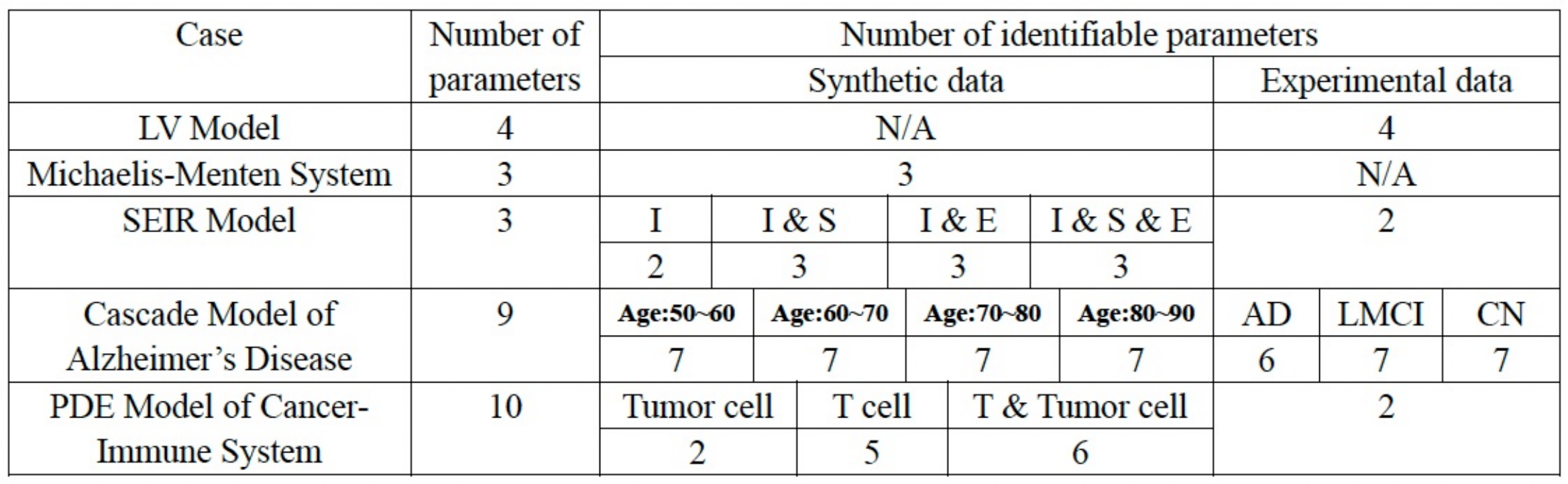}
  
  \label{tab:1}
\end{table}

For cases where the FIM is singular, we approach the problem from two perspectives: introducing regularization terms and refining parameter uncertainty quantification. First, in systems biology, previous studies have utilized regularization techniques, such as Tikhonov regularization or functions derived from prior information, to constrain specific parameters during optimization, effectively preventing changes in the loss function \cite{gabor2015robust,liu2017mathematical,wang2025mathematical}. Expanding on this concept, we propose a novel regularization term based on parameter practical identifiability analysis (Figure \ref{fig:1}). Our approach targets non-identifiable parameters by incorporating regularization terms into the loss function, thereby ensuring that all parameters in the model achieve practical identifiability. Additionally, we provide formal proof that the inclusion of this regularization guarantees the practical identifiability of all parameters (Details in Materials and Methods). Second, traditional methods for uncertainty quantification often involve perturbing all parameters simultaneously. This approach inadvertently modifies the loss function value, making it methodologically inconsistent, as uncertainty originates from non-identifiable parameters alone. To address this limitation, we develop an uncertainty quantification that focuses solely on non-identifiable parameters (Figure \ref{fig:1}). This method (Details in Materials and Methods) enables a more precise assessment of the influence of parameter uncertainty on model predictions. We apply our framework across various biological models, incorporating both synthetic and experimental data, and identify distinct differences in practical identifiability arising from factors such as the selection of observed variables and data sources (See Table \ref{tab:1}). The threshold values for practical identifiability analysis are typically determined in accordance with the context and modeling background of each application (Table S3 in Supplementary Materials).

The integration of mathematical models and data is essential in systems biology, yet determining how models can effectively guide data measurement remains a critical, unresolved challenge. For specific models, it is vital to design optimal experimental data collection strategies grounded in parameter practical identifiability \cite{gevertz2024minimally}. Addressing this challenge, and leveraging our advancements in practical identifiability, we develop an algorithm to generate an optimal sequence of experimental measurement time points. This approach ensures that the collected data render all model parameters practically identifiable. We validate the algorithm by applying it to the Hill function (Figure \ref{fig:3}d) and the Michaelis–Menten system (Figure \ref{fig:6}b), successfully identifying critical data points that constitute the minimal dataset required for render parameter practical identifiability. 

In conclusion, we present a novel framework for practical identifiability analysis, grounded in a rigorous new definition of practical identifiability. The framework systematically integrates the essential properties of practical identifiability and introduces innovative tools, such as novel regularization terms and uncertainty quantification methods. Building on these principles, we develop an algorithm designed to guide optimal data collection, ensuring that experimental data robustly supports model parameter practical identifiability. Our practical identifiability analysis framework demonstrates substantial potential as a crucial bridge between mathematical modeling and experimental data in systems biology. This paper focuses on the least squares loss function and establishes the framework under the assumption of small residuals in model fitting. Under this condition, the FIM provides a good approximation to the Hessian of the loss function. For more general loss functions, such as cross-entropy, or in cases with large residuals under the least squares setting, practical identifiability analysis can be straightforwardly extended by replacing the FIM with the Hessian matrix.

\section*{Materials and Methods}

\subsection*{Practical Identifiability Analysis Using the FIM}
For the time-series data-driven modeling approach, the loss function $l(\boldsymbol{h}(\boldsymbol{\varphi(t, \theta)}), \hat{\boldsymbol{h}})$ is defined using the least squares method as follows:
\begin{equation}
\label{eq:1}
l(\boldsymbol{h}(\boldsymbol{\varphi}(t, \boldsymbol{\theta})), \hat{\boldsymbol{h}}) = \sum_{i=1}^N \| \boldsymbol{h}(\boldsymbol{\varphi}(t_i, \boldsymbol{\theta})) - \hat{\boldsymbol{h}}_i \|_2^2
\end{equation}
where $N$ is the number of experimental data, $\boldsymbol{\varphi}(\boldsymbol{t}, \boldsymbol{\theta})\in \mathbb{R} ^M$ denotes the system output with parameter $\boldsymbol{\theta}$ at the time $\boldsymbol{t}=[t_1,t_2,…,t_N ]^T$ ( $\boldsymbol{\varphi(t, \theta)} = 
\begin{bmatrix}
\varphi_1\boldsymbol{(t, \theta)}, \varphi_2\boldsymbol{(t, \theta)}, \cdots, \varphi_M\boldsymbol{(t, \theta)}
\end{bmatrix}^T $, and $\varphi_i\boldsymbol{(t, \theta)} = 
\begin{bmatrix}
\varphi_i(t_1, \boldsymbol{\theta}), \varphi_i(t_2, \boldsymbol{\theta}), \cdots, \varphi_i(t_N, \boldsymbol{\theta})
\end{bmatrix}^T $). The  experimental observation is denoted as $\{(t_i, \hat{\boldsymbol{h}}_i)\}_{i=1}^N, \quad (\hat{\boldsymbol{h}} = 
\begin{bmatrix}
\hat{\boldsymbol{h}}_1, \hat{\boldsymbol{h}}_2, \dots, \hat{\boldsymbol{h}}_N
\end{bmatrix}^T)$, and the continuous differentiable function $\boldsymbol{h}(\cdot)$ represents measurable quantities ($\boldsymbol{h}(\cdot)\in\mathbb{R}^L$). The parameters of this system $\boldsymbol{\theta^*}$ are given as 
\begin{equation}
\label{eq:2}
\boldsymbol{\theta^*} = \arg\min_{\boldsymbol{\theta} \in \boldsymbol{\Theta}} l(\boldsymbol{h(\varphi(t, \theta)), \hat{h})}
\end{equation}
where $\boldsymbol{\theta}$ is the parameter space. The parameter of this system $\boldsymbol{\theta_\delta}$ for the presence of small perturbation ($\boldsymbol{\delta}$) in measurements is obtained as
\begin{equation}
\label{eq:3}
\boldsymbol{\theta_\delta} = \arg\min_{\boldsymbol{\theta} \in \boldsymbol{\Theta}} l(\boldsymbol{h(\varphi(t, \theta)), \hat{h}-\delta)}
\end{equation}
Herein, the loss function $l(\boldsymbol{h(\varphi(t, \theta)), \hat{h}-\delta)}$
is hypothesized to be continuous with respect to small perturbation ($\boldsymbol{\delta}$). We propose a novel definition of parameter practical identifiability based on the changes in parameters resulting from measurement perturbations (Eqs. \ref{eq:2}-\ref{eq:3}) as follows:

\noindent \textbf{Definition 1}: The parameter $\boldsymbol{\theta}$ in $\boldsymbol{\Theta}$ is practically identifiable if $\forall \varepsilon > 0, \ \exists C > 0 \text{ such that } \|\boldsymbol{\delta}\| < \varepsilon \implies \|\boldsymbol{\theta_\delta} - \boldsymbol{\theta^*}\| < C\varepsilon
$ where $\boldsymbol{\theta^*}$ and $\boldsymbol{\theta_\delta}$ satisfy Eq. \ref{eq:2} and Eq. \ref{eq:3}, respectively.
Then we define the general sensitive matrix $s(\boldsymbol{\theta^*})$ with the observable function $\boldsymbol{h}(\cdot)\in \mathbb{R}^L$ as:
\begin{equation}
\label{eq:4}
\boldsymbol{s(\theta^*)} = 
\begin{bmatrix}
\boldsymbol{s_1}(\boldsymbol{\theta^*});
\boldsymbol{s_2}(\boldsymbol{\theta^*});
\cdots;
\boldsymbol{s_N}(\boldsymbol{\theta^*}) \nonumber
\end{bmatrix}_{N \times 1}, 
\end{equation}
\begin{equation}
\boldsymbol{s_n(\theta^*)} = \nabla_{\boldsymbol{\theta}} \boldsymbol{h} (\boldsymbol{\varphi}(t_n, \boldsymbol{\theta^*}))= 
\begin{bmatrix}
s_{11}(t_n) & s_{12}(t_n) & \cdots & s_{1k}(t_n) \\
s_{21}(t_n) & s_{22}(t_n) & \cdots & s_{2k}(t_n) \\
\vdots & \vdots & \ddots & \vdots \\
s_{L1}(t_n) & s_{L2}(t_n) & \cdots & s_{Lk}(t_n)
\end{bmatrix}_{L \times k} (n=1,2,...,N), \nonumber
\end{equation}
\begin{equation}
s_{li}(t_n) =\frac{\partial h_l(\boldsymbol{\varphi}(t_n,\boldsymbol{\theta^*}))}{\partial \theta_i}= \sum_{m=1}^M \left( \frac{\partial h_l(\boldsymbol{\varphi})}{\partial \varphi_m} \cdot \frac{\partial \varphi_m(t_n, \boldsymbol{\theta^*})}{\partial \theta_i} \right)(l=1,2,...L;i=1,2,...,k).
\end{equation}

Then FIM is defined as follows:
\begin{equation}
\label{eq:5}
F(\boldsymbol{\theta^*})=\frac{1}{\sigma^2}s^T(\boldsymbol{\theta^*})s(\boldsymbol{\theta^*})
\end{equation}
Based on these definitions, we explore the relationship between parameter practical identifiability and FIM as stated in Theorem 1:

\noindent\textbf{Theorem 1}: The parameter $\boldsymbol{\theta}$ in $\boldsymbol{\Theta}$ is practically identifiable if and only if the FIM $F(\boldsymbol{\theta^*})$ is invertible. (Details of the proof in the “Proof of Theorem 1” section in Supplementary Materials)
\subsection*{Coordinate Parameter Identifiability}
Coordinate parameter identifiability is defined using the Bayesian posterior likelihood \cite{wieland2021structural,raue2009structural,raue2013joining} as follows:

\noindent \textbf{Definition 2}: The parameter $\boldsymbol{\theta}$ is coordinate identifiable if the profile likelihood $PL(\boldsymbol{\hat{h}} \mid \theta_i) = \min_{\theta_{j \neq i}} \big[l(\boldsymbol{\hat{h}}; \boldsymbol{\theta})\big]
$ has a locally unique minimum at $\theta_i^*$ for each parameter coordinate $i$.

Considering that the coordinate parameter identifiability is local, we use the linear approximation to investigate the relationship between the practical identifiability and coordinate identifiability at the given parameter point $\boldsymbol{\theta^*}$. First, the observable quantities of the system $\boldsymbol{h(\varphi(t,\theta))}$ at the fixed time $t=t_j$ is linearly approached as:

\begin{equation}
\label{eq:6}
\boldsymbol{h}(\boldsymbol{\varphi}(t_j,\boldsymbol{\theta})) \approx \boldsymbol{h}(\boldsymbol{\varphi}(t_j,\boldsymbol{\theta^*}))+\boldsymbol{s_j(\theta^*)(\theta-\theta^*)}
\end{equation}
where $\boldsymbol{s_j(\theta^*)}$ is defined in Eq. \ref{eq:4}. The logarithmic likelihood function $l(\boldsymbol{\hat{h}; \theta}) \equiv -\log L(\boldsymbol{\hat{h}; \theta})$ is given as
\begin{equation}
\label{eq:7}
\min_{\boldsymbol{\theta}} l(\boldsymbol{\hat{h}; \theta}) \approx \min_{\boldsymbol{\theta}} \| \boldsymbol{h(\varphi(t, \theta^*))} + \boldsymbol{s(\theta^*)(\theta - \theta^*)} - \boldsymbol{\hat{h}} \|_2^2
\end{equation}
where $\boldsymbol{h(\varphi(t, \theta))}= 
\begin{bmatrix}
\boldsymbol{h}(t_1, \boldsymbol{\theta}),  \cdots, \boldsymbol{h}(t_N, \boldsymbol{\theta})
\end{bmatrix}^T $, and $\boldsymbol{h}(\boldsymbol{\varphi}(t_i, \boldsymbol{\theta})) = 
\begin{bmatrix}
h_1(\boldsymbol{\varphi}(t_i, \boldsymbol{\theta}), h_2(\boldsymbol{\varphi}(t_i, \boldsymbol{\theta}), \cdots, h_L(\boldsymbol{\varphi}(t_i, \boldsymbol{\theta})
\end{bmatrix}^T $ is the observable system output with parameter $\boldsymbol{\theta}$ and the experiment data is denoted as $\{t_i,\hat{\boldsymbol{h}}\}_{i=1}^N$ $(\boldsymbol{\hat{h}}= 
\begin{bmatrix}
\boldsymbol{\hat{h}}_1, \boldsymbol{\hat{h}}_2, \cdots, \boldsymbol{\hat{h}}_N
\end{bmatrix}^T )$. We denote the sensitive matrix at the parameter $\boldsymbol{\theta^*}$ as $\boldsymbol{S=s(\theta^*)}$ and the constant vector as $\boldsymbol{b=\hat{h}-h(\varphi(t,\theta^* ))+s(\theta^* ) \theta^*}$ so that Eq. \ref{eq:7} is rewritten as
\begin{equation}
\label{eq:8}
\min_{\boldsymbol{\theta}} l(\boldsymbol{\hat{h}; \theta}) \approx \min_{\boldsymbol{\theta}} \| \boldsymbol{S\theta-b} \|_2^2
\end{equation}

\noindent\textbf{Theorem 2:} The parameter $\boldsymbol{\theta}$ is coordinate identifiable if and only if the FIM $F(\boldsymbol{\theta^*})$ is invertible. (Details of the proof in the “Proof of Theorem 2” section in Supplementary Materials)

If the FIM is singular, we further investigate the coordinate non-identifiability using the sensitive matrix $s(\boldsymbol{\theta^*})$ as follows:

\noindent\textbf{Theorem 3:} The parameter $\theta_i\in \boldsymbol{\theta}$ is non-identifiable if and only if $\boldsymbol{s_i} \in \text{range}(A)$. Here, $\boldsymbol{s_i}$ is the $i^{th}$ column of matrix $s(\boldsymbol{\theta^*})$ and $A=[s_1,s_2,\cdots,s_{i-1},s_k,s_{i+1},\cdots,s_{k-1}]$. In another word, $A$ is the $(k-1)$ column of matrix $\hat{S}=SP_{i,k}$, where $P_{i,k}$=$[e_1,e_2,\cdots,e_{i-1},e_k,e_{i+1},\cdots,e_{k-1},e_i]$  is the elementary matrix and the vector $e_i$ is unit vector. (Details in ‘Proof of Theorem 3’ section in Supplementary Materials)

\subsection*{Parameter Regularization and Uncertainty Quantification Based on Practical Identifiability}
\subsubsection*{Parameter Regularization} Based on Bayes’ theorem, the likelihood function is extended by the prior probability density function (PDF) of the parameter $P(\boldsymbol{\theta})$ and the posterior PDF of the parameters is given as
\begin{equation}
\label{eq:9}
P(\boldsymbol{\theta} \mid \hat{\boldsymbol{h}}) = \frac{P(\hat{\boldsymbol{h}} \mid \boldsymbol{\theta}) P(\boldsymbol{\theta})}{P(\hat{\boldsymbol{h}})}, \quad P(\hat{\boldsymbol{h}} \mid \boldsymbol{\theta}) = L(\hat{\boldsymbol{h}}; \boldsymbol{\theta})
\end{equation}
where $P(\hat{\boldsymbol{h}})$ is the PDF of the experimentally observable data. The parameter $\boldsymbol{\theta^*}$ is obtained by maximum a posteriori (MAP) estimation as
\begin{equation}
\label{eq:9}
\boldsymbol{\theta^*} = \arg\max_{\boldsymbol{\theta} \in \boldsymbol{\Theta}} \log P(\boldsymbol{\theta} \mid \hat{\boldsymbol{h}}) = \arg\min_{\boldsymbol{\theta} \in \boldsymbol{\Theta}} \big(-\log L(\hat{\boldsymbol{h}}; \boldsymbol{\theta}) - \log P(\boldsymbol{\theta})\big)
\end{equation}
Herein, the prior PDF of the parameter $P(\boldsymbol{\theta})$ can be seen as the normalization to the parameter $\boldsymbol{\theta}$. The relative entropy is shown as follows: 
\begin{equation}
\label{eq:11}
\begin{aligned}
D(L(\hat{\boldsymbol{h}}; \boldsymbol{\theta^*}): L(\hat{\boldsymbol{h}} - \boldsymbol{\delta}; \boldsymbol{\theta_\delta})) = \frac{1}{2\sigma^2} \boldsymbol{\delta^T\delta}  + \frac{1}{\sigma^2} \boldsymbol{\delta^T} \nabla \boldsymbol{h(\varphi(t, \theta^*))}  
 - \frac{1}{2} (\boldsymbol{\theta_\delta - \theta^*})^T F_\delta(\boldsymbol{\theta^*})(\boldsymbol{\theta_\delta - \theta^*})
\end{aligned}
\end{equation}
According to the limitation $\lim_{\|\boldsymbol{\delta}\| \to 0} D(L(\hat{\boldsymbol{h}}; \boldsymbol{\theta}) : L(\hat{\boldsymbol{h}} - \boldsymbol{\delta};\boldsymbol{\theta_\delta})) = 0$, we have $(\boldsymbol{\theta_\delta - \theta^*})^TF(\boldsymbol{\theta^*})(\boldsymbol{\theta_\delta - \theta^*})=0$. We perform the eigenvalue decomposition \cite{golub2013matrix} to FIM $F(\boldsymbol{\theta^*})$ as:
\begin{equation}
\label{eq:12}
F(\boldsymbol{\theta^*}) = U \Sigma U^T, \quad \Sigma = 
\begin{bmatrix}
\Lambda_{r \times r} & 0 \\
0 & 0
\end{bmatrix}  \quad U = [U_r, U_{k-r}] 
\end{equation}
where $U^T U = I_k$, and the Eq. \ref{eq:12} is transformed as
\begin{equation}
\label{eq:13}
\left([U_r^T; U_{k-r}^T]\boldsymbol{(\theta_\delta - \theta^*)}\right)^T 
\begin{bmatrix}
\Lambda_{r \times r} & 0 \\
0 & 0
\end{bmatrix}
\left([U_r^T; U_{k-r}^T]\boldsymbol{(\theta_\delta - \theta^*)}\right) = 0
\end{equation}
The Eq. \ref{eq:13} reflects that the $(k-r)$ parameters are non-identifiable and that the $r$ parameters are practical identifiable because of $\lim_{\|\boldsymbol{\delta}\| \to 0} \|U_r^T\boldsymbol{ (\theta_\delta - \theta^*)}\| = 0$. Moreover, the prior PDF of the parameter $P(\boldsymbol{\theta})$ can be assumed as the gauss distribution at the low dimensional space $U_{k-r}^T \boldsymbol{ (\theta_\delta - \theta^*)} \sim \mathcal{N}(0, \Sigma_1), \quad \Sigma_1 = \tau^2 I_{k-r}
$, so that the regularization denoted as $\log  
P(\boldsymbol{\theta})$ of the parameter $\theta_i^*$:
\begin{equation}
\label{eq:14}
\log P(\theta) = \log \left(\frac{1}{(2\pi \tau^2)^{(k-r)/2}}\right) - \frac{1}{2\tau^2} \|U_{k-r}^T \boldsymbol{\theta} - U_{k-r}^T \boldsymbol{\theta^*}\|_2^2
\end{equation}
For the parameter $\boldsymbol{\theta}$, the regularization without constant part is given as $\lambda\|U_{k-r}^T \boldsymbol{\theta} - U_{k-r}^T \boldsymbol{\theta^*}\|_2^2  (\lambda=1/(2\tau^2 ))$. The MAP estimation is rewritten as
\begin{equation}
\label{eq:15}
\tilde{\boldsymbol{\theta}} = \arg\min_{\boldsymbol{\theta} \in \boldsymbol{\Theta}} \big(-\log L(\hat{\boldsymbol{h}}; \boldsymbol{\theta}) + \lambda \|U_{k-r}^T \boldsymbol{\theta} - U_{k-r}^T \boldsymbol{\theta^*}\|_2^2\big)
\end{equation}
$\tilde{\boldsymbol{\theta}}$ is practically identifiable because the necessary condition of Eq. \ref{eq:15} is $(\lambda U_{k-r} U_{k-r}^T + \boldsymbol{S}^T \boldsymbol{S}) \tilde{\boldsymbol{\theta}} = \boldsymbol{S}^T \boldsymbol{b}
$. The FIM of the parameter $\tilde{\boldsymbol{\theta}}$ is
\begin{equation}
\label{eq:16}
F(\tilde{\boldsymbol{\theta}}) = \lambda U_{k-r} U_{k-r}^T + \boldsymbol{S}^T \boldsymbol{S} = U 
\begin{bmatrix}
\Lambda_{r \times r} & 0 \\
0 & \lambda I_{k-r}
\end{bmatrix}
U^T
\end{equation}
$F(\tilde{\boldsymbol{\theta}})$ is full rank, and the parameter $\tilde{\boldsymbol{\theta}}$ is coordinate identifiable according to Theorem 2. 
\subsubsection*{Uncertainty quantification} We propose an uncertainty quantification method based on practical identifiability to examine the impact of variations in the non-identifiable parameters on the model's uncertainty, ensuring that the observations remain within the defined confidence intervals. To address uncertainties in the parameters, especially those aligned with the non-identifiable eigenvectors $U_{k-r}^T$, we perform a perturbation vector as $\varepsilon_{k-r}\sim N(\boldsymbol{0},\Sigma_{k-r})$  ($U_{k-r}^T \hat{\boldsymbol{\theta}} = U_{k-r}^T \tilde{\boldsymbol{\theta}} + \varepsilon_{k-r}$). The model parameters are adjusted by:
\begin{equation}
\hat{\boldsymbol{\theta}} = \tilde{\boldsymbol{\theta}} + U_{k-r} \varepsilon_{k-r} 
\end{equation}
The observable variable $\boldsymbol{h}(\boldsymbol{\varphi}(t,\boldsymbol{\tilde{\theta}})$ is linearly approached as
\begin{equation}
\boldsymbol{h}(\boldsymbol{\varphi}(t,\boldsymbol{\hat{\theta}})=\boldsymbol{h}(\boldsymbol{\varphi}(t,\boldsymbol{\tilde{\theta}})+\nabla_{\boldsymbol{\theta}}\boldsymbol{h}(\boldsymbol{\varphi}(t,\boldsymbol{\tilde{\theta}})(\hat{\boldsymbol{\theta}}-\tilde{\boldsymbol{\theta}})
\end{equation}
Based on law of propagation of uncertainty, the estimation of uncertainty on the observable variable such as $h_l (\boldsymbol{\varphi} (t, \hat{\boldsymbol{\theta}}))(l = 1, 2, \dots, L),\forall t > 0$ is written as:
\begin{equation}
\label{eq:19}
\begin{aligned}
\text{Var}(h_l (\boldsymbol{\varphi}(t, \boldsymbol{\hat{\theta}}))) = \nabla_\theta h_l (\boldsymbol{\varphi}(t, \tilde{\boldsymbol{\theta}})) \, \text{CoV}(\hat{\boldsymbol{\theta}}) \, (\nabla_\theta h_l (\boldsymbol{\varphi}(t, \tilde{\boldsymbol{\theta}})))^T 
\end{aligned}
\end{equation}
where the variance of parameter $\hat{\boldsymbol{\theta}}$ is obtain as
$\text{CoV}(\hat{\theta}) = U_{k-r} \Sigma_{k-r} U_{k-r}^T$. Through the linear approximation, the variance of the state variable is calculated using the error propagation formula, which can then be used to construct the confidence interval for the state variable. Assuming each component of observable variable $h_l (\boldsymbol{\varphi}(t, \hat{\boldsymbol{\theta}})) (l = 1, 2, \dots, L)$ approximately follows a normal distribution, its $100(1-\alpha)\%$ confidence interval follows:
\begin{equation}
\label{eq:20}
\begin{aligned}
h_l (\boldsymbol{\varphi}(t, \hat{\boldsymbol{\theta}})) \in  \left[h_l (\boldsymbol{\varphi}(t, \tilde{\boldsymbol{\theta}})) - z_{\alpha/2} \sqrt{\text{Var}(h_l (\boldsymbol{\varphi}(t, \tilde{\boldsymbol{\theta}})))}, \right. 
 \left. h_l (\boldsymbol{\varphi}(t, \tilde{\boldsymbol{\theta}})) + z_{\alpha/2} \sqrt{\text{Var}(h_l (\boldsymbol{\varphi}(t, \tilde{\boldsymbol{\theta}})))} \right]   
\end{aligned}
\end{equation}
where $z_{\alpha/2}$ is the critical value of the standard normal distribution.
\subsection*{Structural Identifiability vs. Practical Identifiability} The definition of structural identifiability is stated as follows \cite{miao2011identifiability}:

\noindent\textbf{Definition 3:} The parameter $\boldsymbol{\theta}$ in $\boldsymbol{\Theta}$ is structural identifiability if $\exists \delta > 0, \, \forall \boldsymbol{\theta} \in U(\boldsymbol{\theta^*}, \delta)$, the following property holds:
\begin{equation}
\forall t > 0, \, \boldsymbol{h(\varphi(t, \theta))} = \boldsymbol{h(\varphi(t, \theta^*))} \implies \boldsymbol{\theta = \theta^*} 
\end{equation}

\noindent\textbf{Theorem 4:} The parameter $\boldsymbol{\theta}$ in $\boldsymbol{\Theta}$ is structurally identifiable if and only if $\forall \{t_i\}_{i=1}^\infty$, there is a subsequence $\{t_{i_j}\}_{j=1}^M (M=L*N\geq k)$, and $s(\boldsymbol{\theta^*})$ has column full rank. Herein, $s(\boldsymbol{\theta^*})$ is the sensitive matrix to the parameter $\boldsymbol{\theta^*}$ at the sequence $\{t_{i_j}\}_{j=1}^M$ and $k$ is the number of parameters. (Details in ‘Proof of Theorem 4’ section in Supplementary Materials)

\subsection*{Quantification of dataset contribution to parameter practical identifiability} We propose a quantitative metric $(\xi)$ to evaluate the contribution of a dataset to the practical identifiability of model parameters. The index $\xi$ is defined as the ratio of the smallest eigenvalue $(\sigma_{min})$ to the largest eigenvalue of the FIM $(\sigma_{max})$ as follows:
\begin{equation}
\xi=\frac{\sigma_{min}}{\sigma_{max}}
\end{equation}
As the dataset size increases, $\xi$ approaches a steady state, which represents the maximum contribution of the dataset to the practical identifiability of the model parameters.
\subsection*{Optimal Data Collection Design}
\begin{algorithm}[ht]
\caption{Optimal Data Collection Algorithm}
\label{alg:practical_identifiability}

\textbf{Input:} Model $\boldsymbol{\varphi(t, \theta)}$, and observation $\boldsymbol{h}(\cdot)$ \\
\textbf{Output:} Time set $T$

\begin{algorithmic}[1]
\State Choose randomly $q$ time points as an initialized time series $T = \{t_j\}_{j=1}^q$, denote the size of the time series as $m = q$, eigenvalue tolerance $\epsilon$, and maximum iteration number as $M$.
\State Perform the eigenvalue decomposition on the FIM $F(\boldsymbol{\theta}^*) = [U_r, U_{k-r}]  \, 
\begin{bmatrix}
\Lambda_{r \times r} &  \\
 & 0
\end{bmatrix}  \, [U_r, U_{k-r}] ^T$.

\While{the total step below the maximum iteration ($m  < M$) and $r< k$}
    \State Find $t_{m+1} \notin T$ through the optimization:
    \[
    t_{m+1}=\arg\max_t \left\| diag(U_{k-r}^T\tilde{S}^T(\boldsymbol{\theta}^*) \, \tilde{S}(\boldsymbol{\theta}^*)U_{k-r}) \right\|_0
    \]
    \State Update $T = T\cup t_{m+1}$, $m = m+1$,  and $F(\theta^*)=F(\theta^*)+\boldsymbol{s_m}^T(\boldsymbol{\theta^*})\boldsymbol{s_m}(\boldsymbol{\theta^*})$.
\State Perform the eigenvalue decomposition $F(\theta^*) = [U_r, U_{k-r}]  \, 
\begin{bmatrix}
\Lambda_{r \times r} &  \\
 & 0
\end{bmatrix}  \, [U_r, U_{k-r}] ^T$.
\EndWhile
\State \textbf{Return:} Time set $T = \{t_j\}_{j=1}^m$
\end{algorithmic}
\end{algorithm}

We develop an optimization algorithm to determine the minimal number of data points required for practical parameter identifiability. Specifically, the algorithm seeks the minimal number of time points \(m\) such that the Fisher Information Matrix (FIM), \(F(\boldsymbol{\theta}^*)\), computed at \(\{t_j\}_{j=1}^m\), attains maximal rank (up to a numerical tolerance \(\epsilon\)).

Assume we start with an initial set of time points \(\{t_j\}_{j=1}^q\). The corresponding FIM can be expressed as  
\begin{equation}
F(\boldsymbol{\theta}^*) = S^T(\boldsymbol{\theta}^*) \, S(\boldsymbol{\theta}^*) 
= [U_r, U_{k-r}]  \, 
\begin{bmatrix}
\Lambda_{r \times r} &  \\
 & 0
\end{bmatrix}  \, [U_r, U_{k-r}] ^T.
\end{equation}
Adding a new time point \(t_{q+1}\) introduces an additional sensitivity matrix \(\tilde{S}(\boldsymbol{\theta}^*)\),
the updated FIM becomes  
\begin{equation}
F(\boldsymbol{\theta}^*) \leftarrow F(\boldsymbol{\theta}^*) + \tilde{S}^T(\boldsymbol{\theta}^*) \, \tilde{S}(\boldsymbol{\theta}^*).
\end{equation}

To optimally select the next time point \(t_{q+1}\), we maximize the contribution of \(\tilde{S}^T \tilde{S}\) in increasing the rank of the FIM. This is achieved by solving  
\begin{equation}
\max_t \| diag(U_{k-r}^T\tilde{S}^T(\boldsymbol{\theta}^*) \, \tilde{S}(\boldsymbol{\theta}^*)U_{k-r})\|_0,
\end{equation}
where \(\|\cdot\|_0\) counts the number of nonzero contributions toward increasing the rank. This procedure is iteratively applied to select additional data points until the desired rank is achieved, ensuring practical identifiability with the minimal number of measurements.


\medskip
\medskip
\subsection*{Data and code availability}
All relevant data are within the manuscript and its Supporting Information files. The public datasets are used in this study. Source codes and data have been deposited on the GitHub repository \\(https://github.com/WilliamMoriaty/Practical-Identifiability ). 

\subsection*{Acknowledgments}
This research supported by National Institute of General Medical Sciences through grant 1R35GM146894. Figures 5a, 8a, and 9a are created in BioRender (Wang, S. (2025) https://BioRender.com/lk8x9v8; Wang, S. (2025) https://BioRender.com/ux55wf1; Wang, S. (2025) https://BioRender.com/aqztz0s). 

\medskip

%
\bibliographystyle{MSP}
\bibliography{reference}

\begin{thebibliography}{10}
\providecommand{\url}[1]{\texttt{#1}}
\providecommand{\urlprefix}{URL }

\bibitem{alon2019introduction}
U.~Alon,
\newblock \emph{An introduction to systems biology: design principles of biological circuits},
\newblock Chapman and Hall/CRC, \textbf{2019}.

\bibitem{qiao2019network}
L.~Qiao, W.~Zhao, C.~Tang, Q.~Nie, L.~Zhang,
\newblock \emph{Cell systems} \textbf{2019}, \emph{9}, 3 271.

\bibitem{ma2009defining}
W.~Ma, A.~Trusina, H.~El-Samad, W.~A. Lim, C.~Tang,
\newblock \emph{Cell} \textbf{2009}, \emph{138}, 4 760.

\bibitem{aguda2008microrna}
B.~D. Aguda, Y.~Kim, M.~G. Piper-Hunter, A.~Friedman, C.~B. Marsh,
\newblock \emph{Proceedings of the National Academy of Sciences} \textbf{2008}, \emph{105}, 50 19678.

\bibitem{lang2021landscape}
J.~Lang, Q.~Nie, C.~Li,
\newblock \emph{Biophysical Journal} \textbf{2021}, \emph{120}, 20 4484.

\bibitem{su2024hodge}
Z.~Su, Y.~Tong, G.-W. Wei,
\newblock \emph{Journal of chemical information and modeling} \textbf{2024}, \emph{64}, 8 3558.

\bibitem{la2018rna}
G.~La~Manno, R.~Soldatov, A.~Zeisel, E.~Braun, H.~Hochgerner, V.~Petukhov, K.~Lidschreiber, M.~E. Kastriti, P.~L{\"o}nnerberg, A.~Furlan, et~al.,
\newblock \emph{Nature} \textbf{2018}, \emph{560}, 7719 494.

\bibitem{sha2024reconstructing}
Y.~Sha, Y.~Qiu, P.~Zhou, Q.~Nie,
\newblock \emph{Nature Machine Intelligence} \textbf{2024}, \emph{6}, 1 25.

\bibitem{lander2002morphogen}
A.~D. Lander, Q.~Nie, F.~Y. Wan,
\newblock \emph{Developmental cell} \textbf{2002}, \emph{2}, 6 785.

\bibitem{shen2022scaling}
J.~Shen, F.~Liu, C.~Tang,
\newblock \emph{Science Bulletin} \textbf{2022}, \emph{67}, 14 1486.

\bibitem{rodriguez2022concentration}
K.~Rodriguez, A.~Do, B.~Senay-Aras, M.~Perales, M.~Alber, W.~Chen, G.~V. Reddy,
\newblock \emph{Science advances} \textbf{2022}, \emph{8}, 31 eabo6157.

\bibitem{wei2016dual}
N.~Wei, Y.~Mori, E.~G. Tolkacheva,
\newblock \emph{Journal of theoretical biology} \textbf{2016}, \emph{397} 103.

\bibitem{kreger2023myeloid}
J.~Kreger, E.~T. Roussos~Torres, A.~L. MacLean,
\newblock \emph{Cancer immunology research} \textbf{2023}, \emph{11}, 5 614.

\bibitem{anderson2024global}
H.~G. Anderson, G.~P. Takacs, D.~C. Harris, Y.~Kuang, J.~K. Harrison, T.~L. Stepien,
\newblock \emph{Journal of Mathematical Biology} \textbf{2024}, \emph{88}, 1 10.

\bibitem{liao2022mathematical}
K.-L. Liao, K.~D. Watt,
\newblock \emph{Bulletin of Mathematical Biology} \textbf{2022}, \emph{84}, 8 82.

\bibitem{zhou2023dynamical}
Z.~Zhou, D.~Li, Z.~Zhao, S.~Shi, J.~Wu, J.~Li, J.~Zhang, K.~Gui, Y.~Zhang, Q.~Ouyang, et~al.,
\newblock \emph{PLOS Computational Biology} \textbf{2023}, \emph{19}, 9 e1011383.

\bibitem{perelson2002modelling}
A.~S. Perelson,
\newblock \emph{Nature reviews immunology} \textbf{2002}, \emph{2}, 1 28.

\bibitem{eisenberg2011mechanistic}
M.~C. Eisenberg, Y.~Kim, R.~Li, W.~E. Ackerman, D.~A. Kniss, A.~Friedman,
\newblock \emph{Proceedings of the National Academy of Sciences} \textbf{2011}, \emph{108}, 50 20078.

\bibitem{stepien2018traveling}
T.~L. Stepien, E.~M. Rutter, Y.~Kuang,
\newblock \emph{SIAM Journal on Applied Mathematics} \textbf{2018}, \emph{78}, 3 1778.

\bibitem{kim2023role}
Y.~Kim, J.~Lee, C.~Lee, S.~Lawler,
\newblock \emph{Journal of Mathematical Biology} \textbf{2023}, \emph{86}, 1 14.

\bibitem{mirzaei2023investigating}
N.~M. Mirzaei, W.~Hao, L.~Shahriyari,
\newblock \emph{Iscience} \textbf{2023}, \emph{26}, 5.

\bibitem{lai2018modeling}
X.~Lai, A.~Stiff, M.~Duggan, R.~Wesolowski, W.~E. Carson~III, A.~Friedman,
\newblock \emph{Proceedings of the National Academy of Sciences} \textbf{2018}, \emph{115}, 21 5534.

\bibitem{davey2024simulating}
M.~Davey, C.~Puelz, S.~Rossi, M.~A. Smith, D.~R. Wells, G.~M. Sturgeon, W.~P. Segars, J.~P. Vavalle, C.~S. Peskin, B.~E. Griffith,
\newblock \emph{PNAS nexus} \textbf{2024}, \emph{3}, 10 pgae392.

\bibitem{peskin1977numerical}
C.~S. Peskin,
\newblock \emph{Journal of computational physics} \textbf{1977}, \emph{25}, 3 220.

\bibitem{peskin2020cardiac}
C.~S. Peskin, D.~M. McQueen,
\newblock \emph{High-Performance Computing in Biomedical Research} \textbf{2020}, 51--59.

\bibitem{gabor2015robust}
A.~G{\'a}bor, J.~R. Banga,
\newblock \emph{BMC systems biology} \textbf{2015}, \emph{9} 1.

\bibitem{gallo2022lack}
L.~Gallo, M.~Frasca, V.~Latora, G.~Russo,
\newblock \emph{Science advances} \textbf{2022}, \emph{8}, 3 eabg5234.

\bibitem{wieland2021structural}
F.-G. Wieland, A.~L. Hauber, M.~Rosenblatt, C.~T{\"o}nsing, J.~Timmer,
\newblock \emph{Current Opinion in Systems Biology} \textbf{2021}, \emph{25} 60.

\bibitem{miao2011identifiability}
H.~Miao, X.~Xia, A.~S. Perelson, H.~Wu,
\newblock \emph{SIAM review} \textbf{2011}, \emph{53}, 1 3.

\bibitem{villaverde2016structural}
A.~F. Villaverde, A.~Barreiro, A.~Papachristodoulou,
\newblock \emph{PLoS computational biology} \textbf{2016}, \emph{12}, 10 e1005153.

\bibitem{ligon2018genssi}
T.~S. Ligon, F.~Fr{\"o}hlich, O.~T. Chi{\c{s}}, J.~R. Banga, E.~Balsa-Canto, J.~Hasenauer,
\newblock \emph{Bioinformatics} \textbf{2018}, \emph{34}, 8 1421.

\bibitem{hong2019sian}
H.~Hong, A.~Ovchinnikov, G.~Pogudin, C.~Yap,
\newblock \emph{Bioinformatics} \textbf{2019}, \emph{35}, 16 2873.

\bibitem{rey2023benchmarking}
X.~Rey~Barreiro, A.~F. Villaverde,
\newblock \emph{Bioinformatics} \textbf{2023}, \emph{39}, 2 btad065.

\bibitem{monsalve2022analysis}
G.~M. Monsalve-Bravo, B.~A. Lawson, C.~Drovandi, K.~Burrage, K.~S. Brown, C.~M. Baker, S.~A. Vollert, K.~Mengersen, E.~McDonald-Madden, M.~P. Adams,
\newblock \emph{Science advances} \textbf{2022}, \emph{8}, 38 eabm5952.

\bibitem{gevertz2024minimally}
J.~L. Gevertz, I.~Kareva,
\newblock \emph{npj Systems Biology and Applications} \textbf{2024}, \emph{10}, 1 2.

\bibitem{raue2009structural}
A.~Raue, C.~Kreutz, T.~Maiwald, J.~Bachmann, M.~Schilling, U.~Klingm{\"u}ller, J.~Timmer,
\newblock \emph{Bioinformatics} \textbf{2009}, \emph{25}, 15 1923.

\bibitem{raue2013joining}
A.~Raue, C.~Kreutz, F.~J. Theis, J.~Timmer,
\newblock \emph{Philosophical Transactions of the Royal Society A: Mathematical, Physical and Engineering Sciences} \textbf{2013}, \emph{371}, 1984 20110544.

\bibitem{ciocanel2024parameter}
M.-V. Ciocanel, L.~Ding, L.~Mastromatteo, S.~Reichheld, S.~Cabral, K.~Mowry, B.~Sandstede,
\newblock \emph{Bulletin of Mathematical Biology} \textbf{2024}, \emph{86}, 4 36.

\bibitem{komorowski2011sensitivity}
M.~Komorowski, M.~J. Costa, D.~A. Rand, M.~P. Stumpf,
\newblock \emph{Proceedings of the National Academy of Sciences} \textbf{2011}, \emph{108}, 21 8645.

\bibitem{joubert2018determining}
D.~Joubert, J.~Stigter, J.~Molenaar,
\newblock \emph{PLoS One} \textbf{2018}, \emph{13}, 11 e0207334.

\bibitem{murray2002mathematical}
J.~Murray,
\newblock Mathematical biology, vol. i, revised 3rd edition, \textbf{2002}.

\bibitem{howard2009modeling}
P.~Howard,
\newblock \emph{Lecture Notes for Math} \textbf{2009}, \emph{442}.

\bibitem{lei2021systems}
J.~Lei,
\newblock \emph{Systems biology},
\newblock Springer, \textbf{2021}.

\bibitem{higham2008modeling}
D.~J. Higham,
\newblock \emph{SIAM review} \textbf{2008}, \emph{50}, 2 347.

\bibitem{hunter2022understanding}
E.~Hunter, J.~D. Kelleher,
\newblock \emph{Journal of Computational Mathematics and Data Science} \textbf{2022}, \emph{4} 100056.

\bibitem{hao2022optimal}
W.~Hao, S.~Lenhart, J.~R. Petrella,
\newblock \emph{PLoS computational biology} \textbf{2022}, \emph{18}, 9 e1010481.

\bibitem{matzavinos2004mathematical}
A.~Matzavinos, M.~A. Chaplain, V.~A. Kuznetsov,
\newblock \emph{Mathematical Medicine and Biology} \textbf{2004}, \emph{21}, 1 1.

\bibitem{brown2003statistical}
K.~S. Brown, J.~P. Sethna,
\newblock \emph{Physical review E} \textbf{2003}, \emph{68}, 2 021904.

\bibitem{liu2017mathematical}
Y.~Liu, X.~Zou,
\newblock \emph{PLoS Computational Biology} \textbf{2017}, \emph{13}, 9 e1005733.

\bibitem{wang2025mathematical}
S.~Wang, T.~Wang, S.~Wu, L.~Zhang, X.~Zou,
\newblock \emph{SIAM Journal on Applied Mathematics} \textbf{2025}, \emph{85}, 1 50.

\bibitem{golub2013matrix}
G.~H. Golub, C.~F. Van~Loan,
\newblock \emph{Matrix computations},
\newblock JHU press, \textbf{2013}.

\end{thebibliography}

\end{document}